%% file: Paper_4.tex
\documentclass[preprint,showpacs,amsmath,amsfonts,amssymb,aps,showkeys]
              {revtex4}
\usepackage{latexsym,makeidx,ifthen,calc,longtable,graphics}
\usepackage{bm}

\begin{document}

\title{Consistent Histories of Systems and Measurements in Spacetime}

\preprint{Version 2.2}

\author{Ed Seidewitz}
\email{seidewitz@mailaps.org}
\affiliation{14000 Gulliver's Trail, Bowie MD 20720 USA}

\date{31 January 2011}


\begin{abstract}

Traditional interpretations of quantum theory in terms of wave
function collapse are particularly unappealing when considering the
universe as a whole, where there is no clean separation between
classical observer and quantum system and where the description is
inherently relativistic. As an alternative, the consistent histories
approach provides an attractive ``no collapse'' interpretation of
quantum physics. Consistent histories can also be linked to
path-integral formulations that may be readily generalized to the
relativistic case. A previous paper described how, in such a
relativistic spacetime path formalism, the quantum history of the
universe could be considered to be an eignestate of the measurements
made within it. However, two important topics were not addressed in
detail there: a model of measurement processes in the context of
quantum histories in spacetime and a justification for why the
probabilities for each possible cosmological eigenstate should follow
Born's rule. The present paper addresses these topics by showing how
Zurek's concepts of einselection and envariance can be applied in the
context of relativistic spacetime and quantum histories. The result is
a model of systems and subsystems within the universe and their
interaction with each other and their environment.

\end{abstract}

\maketitle


\input{Paper_4_Body}

\bibliography{../../RQMbib}

\end{document}

%% file: Paper_4_Body.tex
\section{Introduction} \label{sect:intro}

In recent years, work on quantum gravity and quantum cosmology, in
particular, has made it impossible to avoid the interpretational
issues of quantum mechanics. When considering the universe as a whole,
there is no longer any clean separation between the observer and the
observed system or between the classical and quantum worlds. Further,
any complete cosmological theory must be fully relativistic. In this
arena, traditional interpretations in terms of wave function collapse
are particularly unappealing.

The consistent or decoherent histories approach provides an attractive
``no collapse'' interpretation of quantum physics
\cite{griffiths84,omnes88,gellmann90,griffiths02}. The basic idea of
the approach is to assign probabilities to \emph{histories}, which are
time-ordered sequences of quantum properties of a system. When a
family of histories are chosen in such a way that they are
\emph{consistent}, they \emph{decohere}, and classical probabilities
can be assigned to them as alternative histories of the system.

Consistent histories can also be linked to path-integral formulations
of quantum mechanics \cite{feynman48,feynman65,caves86}. If the
quantum properties under consideration can be expressed in terms of
particle positions, then a quantum history can be considered to be a
superposition of those paths in which the particle passes through
positions with the required properties at the required times. Looked
at another way, a particle path can be considered to be a
\emph{fine-grained} history in which the particle position is exactly
determined at every time, while a path integral represents a
\emph{coarse-grained} history as a superposition of all paths meeting
some more general criteria. When the criteria are properly chosen, the
states for these coarse-grained histories decohere and can have
classical probabilities assigned to them
\cite{hartle91b,hartle92,hartle95}.

The path-integral formulation can also be readily generalized to the
relativistic case by replacing paths in space parameterized by time
with paths in spacetime parameterized by an invariant path
evolution parameter \cite{hartle95,feynman49,feynman50,feynman51,%
teitelboim82,hartle83,hartle86,halliwell01a,halliwell01b,halliwell02}.
The same consistent histories interpretation of path integrals can
then be carried over from the non-relativistic to the relativistic
case, where ``position'' is now understood to mean four-position in
spacetime. (For other approaches to relativistic generalization of the
consistent histories approach, see
\refcite{blencowe91,isham98,isham01,griffiths02a}.)

In \refcite{seidewitz06b} I presented an approach, based on a
relativistic spacetime path formalism \cite{seidewitz06a,seidewitz09},
in which entire coarse-grained histories of the universe decohere for
all time. Each coarse-grained history of all of spacetime is
represented by a \emph{cosmological state} that is constrained by
correlations introduced by the measurement-like processes that occur
within the history of the universe. The cosmological state is
essentially an eigenstate of the operators representing those
processes, and each such state is orthogonal to the rest.

It is only necessary to consider one of these cosmological states to
be the ``real'' history of the actual universe, though, of course, we
have only very partial information on which history this actually is.
Nevertheless, it is shown in \refcite{seidewitz06b} that, if this
``real'' history is selected with a probability determined by the
normal Born rule for the cosmological states, then, from within the
history, all observations made at the classical level can be expected
to be distributed according to the statistical rules of quantum
theory.

Two important topics were not addressed in \refcite{seidewitz06b},
however. First, no detailed model was given of how a measuring
apparatus, as a part of the universe being measured, becomes
correlated with some other part of the universe and itself decoheres
into non-interfering states. Second, no justification of Born's rule
was given for cosmological states; it was simply shown that the
assumption of Born's rule for cosmological states leads to the proper
statistical distribution for repeated experiment results (avoiding the
circularity problem with some previous arguments based on relative
frequencies \cite{kent90,squires90}).

In this paper, I will address both of these topics (though, for the
present paper, I will take a somewhat restricted view of what a
``subsystem'' is, as described in \sect{sect:systems}). In doing so,
it is important to not implicitly presuppose the results of
\refcite{seidewitz06b}, but, rather, to independently support the
assumptions made there, in order not to re-introduce circularity
problems. As noted in \refcite{seidewitz06b}, the arguments of Zurek
on environment-induced superselection (\emph{einselection}) and
entanglement-assisted invariance (\emph{envariance}) are particularly
relevant in this regard.

Zurek has written extensively on einselection and envariance in the
non-relativistic context (see, for example,
\cite{zurek98,zurek03a,zurek03b,zurek05,zurek07a,zurek07b}). Here, I
extend these concepts to the context of a relativistic, spacetime
formalism for quantum histories. Einselection addresses the first of
the topics introduced above, while envariance addresses the second.

\Sect{sect:histories} provides a brief overview of the the consistent
histories approach to non-relativistic quantum mechanics and motivates
the relativistic generalization developed subsequently. This
generalization is grounded in the spacetime path formalism, but the
mathematics of path integration is not actually required for the
results discussed in this paper. The underlying mathematics can
instead be packaged in the familiar notations of relativistic states
and fields. However, the traditional quantum field theory formalism
still presents some difficulties for a straightforward description of
relativistic quantum histories. \Sect{sect:formalism} addresses these
issues through a modified spacetime formalism. The grounding of this
formalism in the spacetime path approach is discussed in the appendix,
making the connection to previous work in
\refcites{seidewitz06b,seidewitz06a,seidewitz09}.

Once the formalism is established, \sect{sect:systems} describes the
foundational concepts of systems and subsystems used in subsequent
sections. The core of the paper comprises \sect{sect:measurements},
which addresses the topic of measurements, and \sect{sect:born}, which
address the topic of Born's rule. \Sect{sect:cat} then applies these
concepts to the paradigmatic thought experiment of Schr\"odinger's
cat. Finally, \sect{sect:conclusion} presents some concluding remarks
on the assumptions underlying Zurek's envariance arguments in relation
to the spacetime approach discussed here.

Throughout, I will use a spacetime metric signature of $(-+++)$ and
take $\hbar = c = 1$.

\section{Consistent Histories} \label{sect:histories}

The consistent histories approach to non-relativistic quantum
mechanics assigns probabilities to quantum \emph{histories}. Such a
history is a sequence of quantum properties at a succession of times
$t_{0} < t_{1} < \ldots < t_{n}$. At each time $t_{i}$, the properties
of interest are represented by a set of projection operators
$\opP^{\alpha_{i}}_{i}$, where the $\alpha_{i}$ label different
properties the system might have at time $t_{i}$. These operators 
satisfy
\begin{equation} \label{eqn:A1}
    \sum_{\alpha_{i}}\opP^{\alpha_{i}}_{i} = 1
\end{equation}
and
\begin{equation*}
    \opP^{\alpha}_{i}\opP^{\beta}_{i} 
	= \delta_{\alpha\beta}\opP^{\alpha}_{i}
\end{equation*}
Each possible history is then completely identified by a sequence of 
labels 
\begin{equation*}
    \boldsymbol{\alpha} = (\alpha_{1}, \alpha_{2}, \ldots, \alpha_{n})
\end{equation*}
and the set of all such histories is known as a \emph{family} of
histories for the system.

Now consider the time evolution of a non-relativistic quantum system
from an initial state $\ket{\Phi(\tz)}$ under a generically
time-dependent Hamiltonian operator $\opH(t)$:
\begin{equation} \label{eqn:A2}
    \ket{\Phi(t)} = T(t,\tz)\ket{\Phi(\tz)} \,,
\end{equation}
where
\begin{equation*}
    T(t_{f},t_{i}) \equiv \me^{-\mi\int^{t_{f}}_{t_{i}} \dt\,\opH(t)}\,.
\end{equation*}
Insert \eqn{eqn:A1} into the time evolution of \eqn{eqn:A2} at each of
the times $t_{i}$:
\begin{alignat*}{2}
    \ket{\Phi(t_{n})} 
	&= \sum_{\alpha_{n}}\opP^{\alpha_{n}}_{n} T(t_{n},t_{n-1})
	   \sum_{\alpha_{n-1}}\opP^{\alpha_{n-1}}_{n-1}
	   T(t_{n-1},&&t_{n-2}) \cdots \\
	&&&\sum_{\alpha_{1}}\opP^{\alpha_{1}}_{1} T(t_{1},\tz)
	   \ket{\Phi(\tz)} \\
	&= \sum_{\boldsymbol{\alpha}} \opC^{\boldsymbol{\alpha}} \ket{\Phi(\tz)} 
	   \,, &&
\end{alignat*}
where
\begin{equation} \label{eqn:A3}
    \opC^{\boldsymbol{\alpha}} \equiv 
	\opP^{\alpha_{n}}_{n} T(t_{n},t_{n-1})
	\opP^{\alpha_{n-1}}_{n-1} T(t_{n-1},t_{n-2})
	\cdots
	\opP^{\alpha_{1}}_{1} T(t_{1},\tz) \,.
\end {equation}

The operator $\opC^{\boldsymbol{\alpha}}$ defined by \eqn{eqn:A3} is
the \emph{chain} or \emph{class} operator for the history
$\boldsymbol{\alpha}$ and the state
$\opC^{\boldsymbol{\alpha}}\ket{\Phi(\tz)}$ is the \emph{branch} of
the system state $\ket{\Phi(t_{n})}$ associated with the history
$\boldsymbol{\alpha}$. We are interested in families of histories such
that all branches are mutually orthogonal, that is,
\begin{equation} \label{eqn:A4}
    \bra{\Phi(\tz)}\opC^{\boldsymbol{\alpha}'}{}\adj
    \opC^{\boldsymbol{\alpha}}\ket{\Phi(\tz)} =
    \delta_{\boldsymbol{\alpha}'\boldsymbol{\alpha}}p(\boldsymbol{\alpha}) \,.
\end{equation}
These are \emph{consistency} or \emph{decoherence} conditions and a
family of histories that satisfies them is known as a
\emph{consistent} or \emph{decoherent} family. $p(\boldsymbol{\alpha})$
then gives the probability for the history $\boldsymbol{\alpha}$.

The left-hand side of \eqn{eqn:A4} is known as the \emph{decoherence 
functional} and is often written in the form
\begin{equation} \label{eqn:A5}
    D(\boldsymbol{\alpha},\boldsymbol{\alpha}') \equiv 
	\bra{\Phi(\tz)}\opC^{\boldsymbol{\alpha}'}{}\adj
	\opC^{\boldsymbol{\alpha}}\ket{\Phi(\tz)}
	    = \Tr(\opC^{\boldsymbol{\alpha}}\rho_{0}
	      \opC^{\boldsymbol{\alpha}'}{}\adj) \,,
\end{equation}
where
\begin{equation*}
    \rho_{0} \equiv \ket{\Phi(\tz)}\bra{\Phi(\tz)} \,.
\end{equation*}
In this form, the decoherence functional can be generalized to 
density matrices $\rho_{0}$ for non-pure initial states, but this 
generality will not be needed here.

Also, the chain operators as defined by \eqn{eqn:A3} satisfy
\begin{equation} \label{eqn:A6}
    \sum_{\boldsymbol{\alpha}}\opC^{\boldsymbol{\alpha}} = T(t_{n},\tz) \,.
\end{equation}
It is perhaps more common to define chain operators in the form
\begin{equation*}
    \opK^{\boldsymbol{\alpha}} \equiv 
	\opP^{\alpha_{n}}_{n}(t_{n})
	\opP^{\alpha_{n-1}}_{n-1}(t_{n-1}) \cdots 
	\opP^{\alpha_{1}}_{1}(t_{1}) \,,
\end{equation*}
where the $\opP^{\alpha_{i}}_{i}(t_{i})$ are the Heisenberg form of the 
projection operators defined by
\begin{equation*}
    \opP^{\alpha_{i}}_{i}(t_{i}) \equiv
	T(\tz,t_{i})\opP^{\alpha_{i}}_{i}T(t_{i},\tz) \,,
\end{equation*}
so that
\begin{equation*}
    \sum_{\boldsymbol{\alpha}}\opK^{\boldsymbol{\alpha}} = 1 \,.
\end{equation*}
However, the two forms of chain operators are related by
\begin{equation*}
    \opK^{\boldsymbol{\alpha}} = 
	T(\tz,t_{n})\opC^{\boldsymbol{\alpha}} \,,
\end{equation*}
and the extra propagator factor cancels out if $\opK$ is used instead
of $\opC$ in the definition of the decoherence functional,
\eqn{eqn:A5}. Thus, the two forms of chain operators are essentially
equivalent as far as consistent histories are concerned. But the form
satisfying the condition of \eqn{eqn:A6} is more closely analogous to
the form of the similar operators to be defined in the following.

As outlined above, a quantum history applies to a single quantum
system. However, such a system may be considered to have subsystems in
the usual way, by taking the Hilbert space for the system to be the
product space of the Hilbert spaces of the subsystems. Projection
operators in the history for the system may then represent properties
of the the system as a whole or either of its subsystems.

In the interesting cases, of course, the Hamiltonian dynamics for such
a system will result in the entanglement of its subsystems. But
consider the simple situation in which there are two subsystems: an
apparatus $\sysA$ designed to measure a quantum system $\sysB$. This
has the basic structure of systems that will be of interest in the
following,

Suppose that $\sysA$ and $\sysB$ are initially both in their own 
identifiable initial states so that
\begin{equation*}
    \ket{\Phi(\tz)} = 
	\ket{\Phi_{\sysA}(\tz)}\ket{\Phi_{\sysB}(\tz)} \,.
\end{equation*}
Further, suppose that the time development of the system from $\tz$ 
to $t_{1}$ does not affect $\sysA$, so that
\begin{equation*}
    T(t_{1},\tz)\ket{\Phi_{\sysA}(\tz)}\ket{\Phi_{\sysB}(\tz)}
	= \ket{\Phi_{\sysA}(\tz)}\ket{\Phi_{\sysB}(t_{1})} \,,
\end{equation*}
while the time development from $t_{1}$ to $t_{2}$ does not affect 
$\sysB$, so that
\begin{equation*}
    T(t_{2},t_{1})\ket{\Phi_{\sysA}(\tz)}\ket{\Phi_{\sysB}(t_{1})}
	= \ket{\Phi_{\sysA}(t_{2})}\ket{\Phi_{\sysB}(t_{1})} \,.
\end{equation*}
Then the only properties of interest at $t_{1}$ will be those about
$\sysB$, and those of interest at $t_{2}$ will be about $\sysA$. That
is, the chain operators will have the form
\begin{equation} \label{eqn:A7}
    C^{(\alpha\beta)} = 
	\opP^{\alpha}_{\sysA}T(t_{2},t_{1})
	\opP^{\beta}_{\sysB}T(t_{1},\tz) \,,
\end{equation}
where the projection operators $\opP^{\alpha}_{\sysA}$ act only on the
component of the system state for subsystem $\sysA$ while the
$\opP^{\beta}_{\sysB}$ act only on the component for $\sysB$.

Now, for an ideal measurement process, the \emph{dynamics} correlates
the pointer states of the apparatus with the states of the measured
subsystem. That is, if $\alpha$ in \eqn{eqn:A7} enumerates pointer
states of $\sysA$ corresponding to similarly enumerated measured
states $\beta$ of $\sysB$, then the time evolution $T(t_{2},t_{1})$
ensures that $C^{(\alpha\beta)}$ is zero unless $\alpha=\beta$. Note
that this does not change that fact that the $\opP^{\alpha}_{\sysA}$
act \emph{only} on $\sysA$ states and the $\opP^{\beta}_{\sysB}$ act
\emph{only} on $\sysB$ states.

The above simple analysis motivates the following approach for moving 
from non-relativistic to relativistic consistent histories.

Consider the subsystems $\sysA$ and $\sysB$ to occupy physical
three-dimensional volumes within the overall combined system (e.g.,
the physical space occupied by the apparatus $\sysA$, etc.). Then take
the time interval $[t_{1},t_{2}]$ along with the 3-volume for $\sysA$,
forming a \emph{four-dimensional} hypervolume within which all
interesting dynamics happens for $\sysA$. Similarly, take the time
interval $[t_{0},t_{1}]$ with the 3-volume for $\sysB$ to form a
hypervolume of interest for $\sysB$.

Heuristically, what is desired is to recast chain operators of the
form of \eqn{eqn:A7} into a form something like
\begin{equation} \label{eqn:A8}
    C^{(\alpha\beta)} = 
	\opP^{\alpha}_{\sysA} \opG_{\sysA}
	\opP^{\beta}_{\sysB} \opG_{\sysB} \,,
\end{equation}
where the operators $\opP^{\alpha}_{\sysA}$ represent properties of
interest about the hypervolume associated with $\sysA$ and
$\opG_{\sysA}$ represents the dynamical interactions that occur in
that hypervolume, and similarly for $\opP^{\beta}_{\sysB}$,
$\opG_{\sysB}$ and $\sysB$. The point of this is to develop a
spacetime formulation for the chain operator that is manifestly
Lorentz invariant.

The use of spacetime hypervolumes here has some similarity to 
previous analyses of the probabilities for a particle to enter a 
specific spacetime region in timeless quantum theories 
\cite{halliwell01b,halliwell02,halliwell06}. However, in the present 
case there will generally be \emph{many} particles in each 
hypervolume (i.e., the particles that physically make up the 
subsystem in that hypervolume) and these particles will be 
interacting within and across the hypervolumes. Thus, to handle 
multiple, interacting particles in spacetime, we turn to the 
formalisms of quantum field theory. 

\section{Spacetime Formalism} \label{sect:formalism}

A (Heisenberg picture) \emph{quantum field} is an operator-valued
function $\oppsi$ on spacetime satisfying the Klein-Gordon equation
\begin{equation} \label{eqn:B1}
    (\frac{\partial^{2}}{\partial x^{2}} - m^{2})\oppsi(x) = 0 \,.
\end{equation}
The operator $\oppsi(x)$ acts on the Fock space of particle position
states, destroying a particle at position $x$, while its adjoint
$\oppsit(x)$ acts to create a particle at $x$ 
\cite{peskin95,weinberg95,ticciati99}.

Distinguish particle fields $\adv{\oppsi}$ from antiparticle fields 
$\ret{\oppsi}$. Each kind of field has a specific nonzero commutator 
with its adjoint:
\begin{equation} \label{eqn:B2}
    [\advret{\oppsi}(x'),\advret{\oppsit}(x)] = \proparsym(x'-x) \,,
\end{equation}
where
\begin{equation*}
    \proparsym(x'-x) \equiv (2\pi)^{3}\intthree p\,
	\frac{\me^{\mi[\mp\Ep(x^{\prime 0}-x^{0}) + 
	               \threep\cdot(\threexp-\threex)]}}
	     {2\Ep} \,,
\end{equation*}
with $\Ep \equiv \sqrt{\threep^{2}+m^{2}}$. (For simplicity I will 
only consider scalar fields here. The generalization to non-scalar 
fields and fermionic anticommutation rules is straightforward and 
does not substantially effect the following discussion.)

Let $\ket{0}$ be the vacuum state of the Fock space. Then single 
particle and antiparticle position states are given by
\begin{equation} \label{eqn:B2a}
    \ketarx = \advret{\oppsit}(x)\ket{0} \,.
\end{equation}
While these states represent a particle or antiparticle localized at
at a specific position $x$, they are not orthogonal and states for
different positions overlap. This is due to the fact that they are
constrained to be \emph{on-shell} by \eqn{eqn:B1}. Indeed, the
commutation relations from \eqn{eqn:B2} imply that
\begin{equation*}
    \inner{\advret{x'}}{\advret{x}} = \proparsym(x'-x) \,.
\end{equation*}
They do obey a completeness relation, but only over any spacelike
hypersurface $\Sigma$, not over all of spacetime \cite{halliwell01b}:
\begin{equation*}
    \mi\int_{\Sigma} \dthree x\,
	( \ketax \overset{\leftrightarrow}{\partial}_{0}\braax - 
	  \ketrx \overset{\leftrightarrow}{\partial}_{0}\brarx 
	) = 1 \,.
\end{equation*}

These properties make the $\ketarx$ states inconvenient for
constructing projection operators. Instead, one would like to have an
orthogonal basis for spacetime position states, analogous to the
three-dimensional spacial position states familiar from
non-relativistic quantum mechanics. Let $\ketxz$ be such position
states, where
\begin{equation} \label{eqn:B3}
    \innerzz{x'}{x} = \delta^{4}(x'-x) \,.
\end{equation} 
Extend these states to the multiparticle position Fock space by 
defining fields $\oppsiz(x)$ so that
\begin{equation*}
    \ketxz = \oppsitz(x)\ket{0} \,.
\end{equation*}
The commutation rule
\begin{equation*}
    [\oppsiz(x'),\oppsitz(x)] = \delta^{4}(x'-x)
\end{equation*}
then leads to the desired orthogonality relation of \eqn{eqn:B3}.

Of course, the $\oppsiz(x)$ field does not satisfy the Klein-Gordon
equation, \eqn{eqn:B1}, so the $\ketxz$ states are off-shell. However,
the $\oppsiz$ field can be used as a basis for redefining the on-shell
$\advret{\oppsi}$ fields as
\begin{equation*}
    \advret{\oppsi}(x) \equiv 
	\intfour \xz\, \propar \oppsiz (\xz) \,.
\end{equation*}
The fields defined in this way \emph{do} satisfy the Klein-Gordon
equation, but they no longer follow the commutation rule of
\eqn{eqn:B2}. Instead, they follow a similar commutation rule with
$\oppsiz(x)$:
\begin{equation*}
    [\advret{\oppsi}(x'),\oppsitz(x)] = \proparsym(x'-x) \,.
\end{equation*}
Using the new fields to create the states $\ketarx$ as in 
\eqn{eqn:B2a} then implies that
\begin{equation} \label{eqn:B3a}
    \innerz{\advret{x'}}{x} = \proparsym(x'-x) \,.
\end{equation}

Actually, more useful in the following will be the field operator
\begin{equation*}
    \oppsi(x) \equiv \intfour \xz\, \prop \oppsiz(\xz) \,,
\end{equation*}
where $\prop$ is the Feynman propagator
\begin{equation*}
    \prop \equiv -\mi (2\pi)^{-4} \intfour p\, 
	\frac{\me^{\mi p\cdot(x-\xz)}}
	     {p^{2}+m^{2}-\mi\varepsilon} \,,
\end{equation*}
with the commutation relationship 
\begin{equation*}
    [\oppsi(x'), \oppsitz(x)] = \propsym(x'-x) \,.
\end{equation*}
and the corresponding position states
\begin{equation*}
    \ketx = \oppsit(x)\ket{0} = \intfour \xz\, \prop \ketz{\xz}
\end{equation*}
such that
\begin{equation} \label{eqn:B3b}
    \innerz{x'}{x} = \propsym(x'-x) \,.
\end{equation}
The $\ketx$ states are again off-shell. However, using the well-known
relation \cite{peskin95,weinberg95,ticciati99}
\begin{equation*}
    \prop = \thetaax\propa + \thetarx\propr \,
\end{equation*}
in \eqn{eqn:B3b}, along with \eqn{eqn:B3a}, gives
\begin{equation} \label{eqn:B3c}
    \innerz{x'}{x} = 
	\theta(x^{\prime 0}-x^{0})\innerz{\adv{x'}}{x} + 
	\theta(x^{0}-x^{\prime 0})\innerz{\ret{x'}}{x} \,.
\end{equation}

In a path integral approach, the probability amplitude
$\innerz{x'}{x}$ can be interpreted as the superposition of the
probability amplitudes for each possible spacetime path from $x$ to
$x'$. One can think of the $\ketz{x}$ states as representing the
position $x$ at which the paths \emph{start}, while the $\ket{x'}$
states represent the position $x'$ at which the paths \emph{end}. The
difference between the states reflects the directionality of the
paths---propagation is always \emph{from} the start of the path
\emph{to} the end of the path. (The appendix covers in more detail the
underlying spacetime path derivation of the formalism used in the main
body of the text.)

The decomposition of \eqn{eqn:B3c} separates the case in which $x'$ is
in the future of $x$ from that in which $x'$ is in the past of $x$.
This shows that, while normal particles propagate into the future,
antiparticles can effectively be considered to propagate
\emph{backwards} in time, into the past
\cite{stueckelberg41,stueckelberg42,feynman49}. This division into
particle and antiparticle paths depends, of course, on the choice of a
specific coordinate system in which to define the time coordinate.
However, if we take the time limit of the end point of the path to
infinity for particles and negative infinity for antiparticles, then
the particle/antiparticle distinction will be coordinate system
independent \cite{seidewitz06a}.

The off-shell states $\ketx$ are particularly useful because 
they essentially represent \emph{virtual} particles and the 
probability amplitudes $\innerz{x'}{x}$ are the propagation 
amplitudes for these particles on the inner edges of Feynman 
diagrams. They can therefore be used to construct a convenient 
representation for the scattering amplitudes of interacting particles.

To do this, it is first necessary to consider multiple fields
corresponding to different \emph{types} of particles that may
interact with each other. Then, an individual interaction vertex can 
be considered an event at which some number of incoming particles are 
destroyed and some number of outgoing particles are created. Note that 
the qualifiers ``incoming'' and ``outgoing'' are being used here in 
the sense of particle paths in spacetime, not in the sense of 
time---that is, the position states $\ketx$ are \emph{not} being 
separated into particle and antiparticle states.

Such an interaction can be modeled using a \emph{vertex operator}
constructed from the appropriate number of annihilation and creation
operators. For example, consider the case of an interaction with two
incoming particles, one of type $a$ and one of type $b$, and
two outgoing particles of the same types. The vertex operator for this
interaction is
\begin{equation*}
    \opV \equiv g \intfour x\,
        \oppsiz^{(a)\dag}(x)\oppsiz^{(b)\dag}(x)
        \oppsi^{(a)}(x)\oppsi^{(b)}(x)
\end{equation*}
where the coefficient $g$ represents the relative probability
amplitude of the interaction.

In the following, it will be convenient to use the special adjoint
$\oppsi\dadj$ defined by
\begin{equation*}
    \oppsi\dadj(x) = \oppsitz(x) \text{ and }
    \oppsiz\dadj(x) = \oppsit(x) \,.
\end{equation*}
With this notation, the expression for $\opV$ becomes
\begin{equation}  \label{eqn:B4}
    \opV = g \intfour x\,
             \oppsi^{(a)\ddag}(x)\oppsi^{(b)\ddag}(x)
             \oppsi^{(a)}(x)\oppsi^{(b)}(x) \,.
\end{equation}

To account for the possibility of any number of interactions, we just 
need to sum up powers of $\opV$ to obtain the \emph{interaction
operator}
\begin{equation} \label{eqn:B5}
    \opG \equiv \sum_{m=0}^{\infty} \frac{(-\mi)^{m}}{m!}\opV^{m}
              = \me^{-i\opV} \,,
\end{equation}
where the $1/m!$ factor accounts for all possible permutations of the
$m$ identical factors of $\opV$. The $-\mi$ factors are introduced so
that $\opG$ is unitary relative to the special adjoint (that is,
$\opG\dadj\opG = \opG\opG\dadj = 1$), so long as $\opV$ is
self-adjoint relative to it (that is, $\opV\dadj = \opV$). 

The self-adjointness of $\opV$ implies that an interaction must have
the same number of incoming and outgoing particles, of the same types,
at least when only one possible type of interaction is involved (as is
the case with the example of \eqn{eqn:B4}). The formalism can be
easily extended to allow for multiple types of interactions by adding
additional terms to the definition of $\opV$. In this case, only the
overall operator $\opV$ needs to be self-adjoint, not the individual
interaction terms.

As mentioned above, normal particle states are obtained in the
$+\infty$ time limit, while antiparticle states are obtained in the
$-\infty$ time limit. Moving to a momentum representation then results
in (multiparticle) on-shell scattering in and out states. These states
can be used with the interaction operator $\opG$ to compute multipoint
interaction amplitudes. Expanding $\opG$ as in \eqn{eqn:B5} gives a
sum of Feynman diagrams for each possible number of interactions. The
time-limit momentum states give the correct amplitudes for the
truncated external legs of the diagrams \cite{seidewitz06a}.

Unfortunately, an interaction operator $\opG$ of the form given in
\eqn{eqn:B5} with a vertex operator $\opV$ of the form shown in
\eqn{eqn:B4} cannot actually generate all Feynman diagrams. For
example, the vertex operator $\opV$ from \eqn{eqn:B4} necessarily has
$\opV\ket{0} = 0$. This means that $\opG$ cannot generate vacuum
bubble diagrams. Indeed, in general, $\opV$ cannot construct a vertex
unless all incoming particles already exist in the incoming state,
which prevents the construction of loops involving particles incoming
from vertexes constructed ``later''.

The problem is that the directionality of propagation implied by 
$\innerz{x'}{x}$ is essentially arbitrary. We could just as well have 
defined ``reverse'' particle states $\ket{\bar{x}}$ and 
$\ketz{\bar{x}}$ such that
\begin{equation*}
    \innerz{\bar{x}'}{\bar{x}} = \innerz{x}{x'} = \propsym(x-x') \,.
\end{equation*}
That is, in terms of the spacetime paths, $\ket{\bar{x}}$ now
represents the \emph{start} of the paths, while $\ketz{\bar{x}}$ 
represents the \emph{end}.

To properly construct all possible Feynman diagrams, it is necessary
to include such reverse particle states. This can be easily done in
the formalism by doubling the Fock space through the addition of a
reverse particle type $\nbar$ corresponding to each particle type $n$,
whose field operators have the commutation rule
\begin{equation*}
    [\oppsi^{(\nbar)}(x'), \oppsi^{(\nbar)\ddag}(x)]
        = \propsym(x-x') \,.
\end{equation*}

Then, define
\begin{equation*}
    \oppsi^{(n')}(x) \equiv 
	\oppsi^{(n)}(x) + \oppsi^{(\nbar)\ddag}(x) \,.
\end{equation*}
That is, using this operator, the destruction of an $n$ particle is
treated as equivalent to the creation of an $\nbar$ particle (and vice
versa). This is similar to the way particle destruction is related to
antiparticle creation in the traditional field theory formalism, but
the distinction for reverse particles is based on direction of
propagation along particle paths (which is Lorentz invariant), not the
direction of propagation in time.

Now use the new operators $\oppsi^{(n')}$ instead of $\oppsi^{(n)}$ in
the construction of the vertex operator $\opV$. For example, an
interaction of the form given in \eqn{eqn:B4} becomes
\begin{equation*}
    \opV = \intfour x\, \opV(x) \,,
\end{equation*}
where
\begin{equation} \label{eqn:B6}
    \opV(x) \equiv g :\oppsi^{(a')\ddag}(x)
		      \oppsi^{(b')\ddag}(x)
		      \oppsi^{(a')\prime}(x)
		      \oppsi^{(b')\prime}(x):
\end{equation}
with $:\cdots:$ representing normal ordering, that is, placing all
$\oppsi\dadj$ operators to the left of all $\oppsi$
operators in any product. The interaction operator $\opG =
\exp(-\mi\opV)$ in this vertex operator will then properly
include all loops. Since the $\nbar$ reverse particle types are only
included in the formalism for this purpose, physical states, such as
in and out scattering states, can be constructed using only the
original particle types (that is, constructed using the original
creation operators $\oppsi^{(n)\ddag}$).

The need to include reverse propagation does complicate a bit the
spacetime formalism for interaction presented here. However, the great
advantage of the result is that \eqn{eqn:B5} defining $\opG$ does not
involve time-ordering, as is found in the Dyson expansion of the usual
scattering operator $\opS$ \cite{peskin95,weinberg95,ticciati99}. This
will be critical when considering the decomposition of the
interactions within a system into a number of distinct subsystems, a
topic to which we turn next.

\section{Systems and Subsystems} \label{sect:systems}

As reiterated by Zurek, ``The Universe consists of systems'' 
\cite{zurek07b}. Quantum processes, measuring apparatuses and 
observers are all systems, and all subsystems of the system that is 
the universe as a whole. We make no \emph{a priori} distinction 
between ``quantum'' systems and ``classical'' systems.

We do, however, need to have a crisp definition of how to delineate
what the systems of interest \emph{are} in any given analysis of some
or all of the universe. As discussed in \sect{sect:histories}, it will
be convenient here to define a \emph{system} as being contained in a
well-defined hypervolume $\sysV$ of spacetime, disjoint from the
hypervolumes occupied by all other systems of interest. This
hypervolume does not have to be continuous or connected, though it
must have measure greater than zero. 

The effective identification of systems and subsystems with regions of
spacetime is somewhat more restrictive than the typical generic
definition used in quantum mechanics, in which a system is taken to be
any portion of the universe whose state can be represented by a vector
in an appropriate Hilbert space. For example, identifying subsystems
with hypervolumes does not allow an overall set of degrees of freedom
of a system to be divided between two subsystems occupying the same
physical space (such as when treating the microscope states of the
molecules of a gas as the ``environment'' for the decoherence of the
aggregate macroscopic properties of the gas). However, the
identification works remarkably well for the situations considered in
this paper, and it should be possible to subsequently extend the
conclusions made here to more abstract concepts of systems.

The main benefit of identifying a system with a specific hypervolume
is that it allows a straightforward definition of what interactions
take place ``within'' the system. Define the \emph{restricted}
interaction operator
\begin{equation*}
    \opV_{\sysV} \equiv \int_{\sysV}\dfour x\, \opV(x) \,,
\end{equation*}
with $\opV(x)$ being the vertex operator at position $x$ (as in
\eqn{eqn:B6}). Then
\begin{equation*}
    \opG_{\sysV} \equiv \me^{-\mi\opV_{\sysV}}
\end{equation*}
generates interactions only within the hypervolume $\sysV$.

Let $\bar{\sysV}$ be the hypervolume of all spacetime with $\sysV$
removed. Define in and out states $\ket{\Psi\In}_{0}$ and
$\ket{\Psi\Out}$ on $\bar{\sysV}$, that is, as superpositions of
states with positions only within $\bar{\sysV}$ and thus
\emph{outside} $\sysV$. Then $\bra{\Psi\Out} \opG_{\sysV}
\ket{\Psi\In}_{0}$ is the scattering amplitude for particles to enter
$\sysV$ (from the in state in $\bar{\sysV}$), interact only within
$\sysV$ and then leave $\sysV$ (to the out state in $\bar{\sysV}$).
(Note that, while all interaction vertices in this process are
restricted to be within $\sysV$, the paths of particles between such
vertices are \emph{not} so restricted.)

Suppose the system contained in $\sysV$ is now further divided into a
set of $N$ \emph{subsystems} with corresponding disjoint hypervolumes
$\sysV_{i}$ such that $\sysV = \bigcup_{i}\sysV_{i}$. Then
\begin{equation} \label{eqn:D1}
    \opG_{\sysV} = \exp\left(-\mi\sum_{i=1}^{N}\opV_{\sysV_{i}}
                       \right)
		 = \prod_{i=1}^{N}\me^{-\mi\opV_{\sysV_{i}}}
		 = \prod_{i=1}^{N}\opG_{\sysV_{i}} \,.
\end{equation}
Of course, the second equality above requires that all the
$\opV_{\sysV_{i}}$ commute. This makes sense conceptually, since the
ordering of the subsystems should not effect the generation of the
full set of interactions for the complete system. It can also be shown
by explicit calculation that $\opV(x_{1})$ and $\opV(x_{2})$ commute
for $x_{1} \neq x_{2}$, so that $\opV_{\sysV_{i}}$ and
$\opV_{\sysV_{j}}$ commute for disjoint $\sysV_{i}$ and $\sysV_{j}$.

Note that the easy decomposition of the $\opG$ operator given in
\eqn{eqn:D1} is only possible because its definition does not require
the time-ordering of interactions, as is embodied in the usual
scattering operator $\opS$. Indeed, the $\opS$ operator may only be
decomposed into commuting factors in the limiting case of widely
separated, non-interacting clusters of particles (the so-called
``cluster decomposition principle'' \cite{weinberg95,wichmann63}),
which is not very useful for the analysis of interacting subsystems.
But note also that the commutivity of the interaction operators
$\opV(x)$ necessary for \eqn{eqn:D1} requires the inclusion of reverse
particle types (as discussed in \sect{sect:formalism}).

We can convert the product of operators in \eqn{eqn:D1} into a 
product of matrix elements using the usual trick of inserting 
resolutions of the identity. For the extended Fock space, such 
resolutions have the form
\begin{equation} \label{eqn:D2}
    1 = \int \dif\chi \dif\bar{\chi}\, \ketz{\chi,\bar{\chi}}
	\braz{\chi,\bar{\chi}} \,, 
\end{equation}
where $\chi,\bar{\chi}$ represent complete configurations of particles
($\chi$) and reverse particles ($\bar{\chi}$). That is,
$\ketz{\chi,\bar{\chi}}$ is a position state of the entire universe,
with $\chi$ symbolically representing the positions of all particles
of normal particle types and $\bar{\chi}$ representing the positions
of all particles of reverse particle types. (For particles of non-zero
spin, appropriate spin indices should also be considered included.)
The integration measure $\int \dif\chi \dif\bar{\chi}$ is intended to
not only include integration of particle positions over all spacetime,
but also summation over all configurations with all possible numbers
of particles of each type (and summation over spin indices, as
appropriate).

Inserting the resolution of the identity from \eqn{eqn:D2} between 
the operators in \eqn{eqn:D1} then gives:
\begin{equation} \label{eqn:D3}
    \opG_{\sysV}
	= \left[
	       \prod_{i=0}^{N}
	       \int \dif\chi_{i} \dif\bar{\chi}_{i}
	   \right]
	   \ketz{\chi_{N},\bar{\chi}_{N}}
	   \left[
	       \prod_{i=1}^{N}
	       \braz{\chi_{i},\bar{\chi}_{i}}\opG_{\sysV_{i}}
	       \ketz{\chi_{i-1},\bar{\chi}_{i-1}}
	   \right]
	   \braz{\chi_{0},\bar{\chi}_{0}} \,.
\end{equation}
Each $\braz{\chi_{i},\bar{\chi}_{i}} \opG_{\sysV_{i}}
\ketz{\chi_{i-1},\bar{\chi}_{i-1}}$ factor is the scattering amplitude
for the interaction of subsystem $\sysV_{i}$ with the rest of the
universe. The interfaces on each ``side'' are \emph{bidirectional}: on
the $i-1$ side, the particles given by $\chi_{i-1}$ are incoming into
$\sysV_{i}$, but the reverse particles in $\bar{\chi}_{i-1}$ are
effectively \emph{outgoing} from $\sysV_{i}$. Similarly, on the $i$
side, the particles in $\chi_{i}$ are outgoing, but the reverse
particles in $\bar{\chi}_{i}$ are effectively \emph{incoming}. Note
again that ``incoming'' and ``outgoing'' are used in the sense of the
propagation along a particle path, not time---incoming particles are
thus not necessarily ``initial'' and outgoing particles are not
necessarily ``final''.

It is important that the integrations in the resolution of the
identity in \eqn{eqn:D2}, and hence also in \eqn{eqn:D3}, cover all of
spacetime, not just the hypervolume of any one system or subsystem.
The expansion of each interaction operator $\opG_{\sysV_{i}}$ includes
the identity operator and hence the possibility of particles passing
through $\sysV_{i}$ (from either ``side'') with no interactions. As a
result, in the absence of any other restrictions, it is possible for a
particle created in an interaction in any of the subsystems (or, for
that matter, outside the system all together) to be annihilated in an
interaction in any of the other subsystems, not just the ones on
either ``side''. This reflects the commutativity of the operators
$\opG_{\sysV_{i}}$, such that the actual ordering of the subsystems is
irrelevant.

The resolution of the identity in \eqn{eqn:D2} can be considered a
superposition of the \emph{most fine-grained} position projection
operators $\ketz{\chi,\bar{\chi}}\braz{\chi,\bar{\chi}}$ that assert
that the universe is in the configuration $\chi,\bar{\chi}$. It is, of
course, generally more useful to make more \emph{course-grained}
assertions about the state of the universe. Such an assertion can be
denoted by a general projection operator $\opP$ on the extended Fock
space.

In particular, it is generally useful to define a \emph{complete} set 
of projection operators $\opP^{\alpha}$ such that
\begin{equation} \label{eqn:D4}
    \sum_{\alpha}\opP^{\alpha} = 1 
\end{equation}
and
\begin{equation*}
    \opP^{\alpha}\opP^{\beta} = \delta_{\alpha\beta}\opP^{\alpha} \,,
\end{equation*}
where, for simplicity, we consider the cardinality of the set to be
finite, or, at least, countable. It should be kept in mind that such
operators define propositions on \emph{all of spacetime}. Some care
needs to be taken when considering propositions on specific systems
within limited hypervolumes of spacetime. 

Given static Minkowski spacetime, an identified hypervolume $\sysV$
will always exist. However, there will be many configurations of the
universe in which $\sysV$ will not actually contain anything we are
interested in---that is, in these configurations the system of
interest presumed to be contained in $\sysV$ will essentially
\emph{not exist}. For example, suppose that the system (or subsystem)
of interest is a measuring instrument, and the projection operators in
\eqn{eqn:D4} represent pointer states of that instrument. But this
presumes that the measuring instrument \emph{is actually there} (and
turn on and operating, etc.) in the expected hypervolume $\sysV$.
There will be many possible configurations of the universe in which
this is just not the case.

Take the projection operator $\opP_{\sysV}^{0}$ to select all
configurations in which a system of interest does \emph{not} exist in
the hypervolume $\sysV$. The complementary operator
$\op{\bar{P}}_{\sysV}^{0}$ then asserts that the system exists in
$\sysV$---and presumably has some additional interesting finer-grained
states, such that
\begin{equation*}
    \opP_{\sysV}^{0} + \op{\bar{P}}_{\sysV}^{0} = 1
\end{equation*}
and
\begin{equation*}
    \op{\bar{P}}_{\sysV}^{0} = \sum_{\alpha>0}\opP_{\sysV}^{\alpha} \,,
\end{equation*}
where the projection operators for actually interesting system states
are denoted by $\opP_{\sysV}^{\alpha}$ for $\alpha>0$. By definition, 
the non-existence assertion $\opP_{\sysV}^{0}$ is never truly 
interesting (at least for the cases considered here), but it is only 
when this operator is included that the set of 
$\opP_{\sysV}^{\alpha}$ is actually complete and sums up to the 
identity as in \eqn{eqn:D4}.

Now, given a system contained in $\sysV$ divided into $N$ subsystems
contained in $\sysV_{i}$, for $i=1,\ldots,N$, suppose we define a
complete set of projection operators $\opP^{\alpha}_{\sysV_{i}}$
corresponding to \emph{each} of the subsystems, as well as a set
$\opP^{\alpha}\In$ defined on the in state for the system. Inserting
\eqn{eqn:D4} for these operators between the scattering operators for
each subsystem in \eqn{eqn:D1} gives
\begin{equation} \label{eqn:D4a}
    \begin{split}
	\opG_{\sysV}
	    &= \left(
	           \sum_{\alpha_{N}}\opP^{\alpha_{N}}_{\sysV_{N}}
	       \right)
	       \opG_{\sysV_{N}}
	       \left(
	           \sum_{\alpha_{N-1}}\opP^{\alpha_{N-1}}_{\sysV_{N-1}}
	       \right)
	       \cdots
	       \left(
	           \sum_{\alpha_{1}}\opP^{\alpha_{1}}_{\sysV_{1}}
	       \right)
	       \opG_{\sysV_{1}}
	       \left(
	           \sum_{\alpha_{0}}\opP^{\alpha_{0}}\In
	       \right) \\
	    &= \sum_{\boldsymbol{\alpha}} \opC^{\boldsymbol{\alpha}} \,,
    \end{split}
\end{equation}
where
\begin{equation}
    \opC^{\boldsymbol{\alpha}} =	    
	\opP^{\alpha_{N}}_{\sysV_{N}}\opG_{\sysV_{N}}
	\opP^{\alpha_{N-1}}_{\sysV_{N-1}}\cdots
	\opP^{\alpha_{1}}_{\sysV_{1}}\opG_{\sysV_{1}}
	\opP^{\alpha_{0}}\In \,,
\end{equation}
for $\boldsymbol{\alpha} = (\alpha_{0},\cdots, \alpha_{N})$. This
equation now has the essential form of \eqn{eqn:A8} desired for chain
operators at the end of \sect{sect:histories} and \eqn{eqn:D4a} is
analogous to the summation of \eqn{eqn:A6} in the non-relativistic
case.

The intent here is that the projection operators
$\opP^{\alpha_{i}}_{\sysV_{i}}$ represent the propositions that either
subsystem $i$ does not exist (represented by $\opP^{0}_{\sysV_{i}}$)
or that there is some ``interesting'' \emph{outcome} $\alpha_{i}$ as a
result of the interactions generated by $\opG_{\sysV_{i}}$
(represented by $\opP^{\alpha_{i}}_{\sysV_{i}}$ for $\alpha_{i}>0$).
We will thus always assume in the following that the assertion made by
each $\opP^{\alpha_{i}}_{\sysV_{i}}$, for $\alpha_{i}>0$), depends
only on outgoing particles from $\sysV_{i}$ that have actually
interacted within $\sysV_{i}$. That is, for all $i$ and any
$\alpha_{i}>0$,
\begin{equation} \label{eqn:D4b}
    \opP^{\alpha_{i}}_{\sysV_{i}}\ketz{\chi,\bar{\chi}}
	= \opP^{\alpha_{i}}_{\sysV_{i}}\ketz{\chi',\bar{\chi}'} \,,
\end{equation}
for all $\bar{\chi}$ and $\bar{\chi}'$ and any $\chi$ and
$\chi'$ that differ only in positions outside of $\sysV_{i}$. (Note
that this assumption now presumes a specific ordering of the
subsystems, but that will actually be convenient in the following, as 
further discussed in \sect{sect:conclusion}.)

Given this assumption, each $\opP^{\alpha_{i}}_{\sysV_{i}}
\opG_{\sysV_{i}} \opP^{\alpha_{i-1}}_{\sysV_{i-1}}$ effectively
defines the probability amplitude for the outcome $\alpha_{i}$ for
subsystem $i$ given the outcome $\alpha_{i-1}$ for subsystem $i-1$.
This cannot exactly be called a ``transition'' amplitude, since there
is not necessarily any ordering in time. Nevertheless, subsystem
outcomes of interest will generally be chosen such that certain
outcomes of subsystem $i-1$ preclude the interactions required to
generate certain other outcomes of subsystem $i$, in which case
$\opP^{\alpha_{i}}_{\sysV_{i}} \opG_{\sysV_{i}}
\opP^{\alpha_{i-1}}_{\sysV_{i-1}}$ will be identically zero. Indeed,
in the following it is usually the case that outcome $\alpha_{i-1}$
for subsystem $i-1$ implies a \emph{specific} outcome
$A_{i}(\alpha_{i-1})$ for subsystem $i$. That is
\begin{equation} \label{eqn:D5}
    \opP^{\alpha_{i}}_{\sysV_{i}} \opG_{\sysV_{i}}
    \opP^{\alpha_{i-1}}_{\sysV_{i-1}} =
	\delta_{\alpha_{i},A_{i}(\alpha_{i-1})}
	\opP^{A_{i}(\alpha_{i-1})}_{\sysV_{i}} \opG_{\sysV_{i}}
	\opP^{\alpha_{i-1}}_{\sysV_{i-1}}
\end{equation}

The situation represented in \eqn{eqn:D5} is essentially a dynamic
process, in which the cross-correlation terms
$\opP^{\alpha_{i}}_{\sysV_{i}} \opG_{\sysV_{i}}
\opP^{\alpha_{i-1}}_{\sysV_{i-1}}$ for $\alpha_{i} \neq
A_{i}(\alpha_{i-1})$ are driven to zero. (Such a process may actually
only take the cross-correlations to approximately zero, not
identically zero, but I will always assume they are zero in the
following.) The dynamics in time are not explicit, of course, in the
spacetime formalism used here. However, the hypervolume $\sysV_{i}$
for any subsystem will generally have a finite temporal extent and it
can be arranged for the projection operators $\opP_{i}$ to represent
states at the upper bound of that extent. Thus, when looked at from a
time-evolution viewpoint, various processes may be taking place within
the hypervolume $\sysV_{i}$ leading to zero cross-correlations for
outgoing particle states.

We will also assume in the following that either all subsystems of
interest ``exist'', or none do. That is,
$\opP^{\alpha_{N}}_{\sysV_{N}} \opG_{\sysV_{N}}
\opP^{\alpha_{N-1}}_{\sysV_{N-1}} \cdots \opP^{\alpha_{1}}_{\sysV_{1}}
\opG_{\sysV_{1}} \opP^{\alpha_{0}}\In$ is identically zero if some,
but not all, of the $\alpha_{i}$ are zero. Which of the terms for
other values of the $\alpha_{i}$ are non-zero depends on the specific
physical situation under consideration. Clearly, with a full theory of
interactions, all such dependencies should be determinable from first
principles. However, for the purposes of the following sections, it
will be enough to simply assert the physical dependencies between
subsystems required in each situation considered.

\section{Measurements} \label{sect:measurements}

To model a measurement process, first consider a hypervolume $\sysV$
bounded by hyperplanes at $t=T\I$ and $t=T\F>T\I$. Given these
temporal bounds, we can reasonably define a truly \emph{initial} state
$\ketz{\Psi\I}$ as a superposition of position states for which all
positions have $t<T\I$. Similarly, define a final state $\ket{\Psi\F}$
as a superposition of position states for which all positions have
$t>T\F$.

Per the discussion in \sect{sect:formalism}, in and out states such as
$\ketz{\Psi\I}$ and $\ket{\Psi\F}$ are required to have no particles
of reverse types. Further, the temporal bounding of $\sysV$ means
that, by construction, in the frame of the time $t$, all initial and
final particles are regular particles, not antiparticles.

Now divide $\sysV$ into two subsystems: a \emph{measured system}
$\sysS$ and a \emph{measuring apparatus} $\sysA$. Define a
complete set of projection operators $\opP^{\alpha}_{\sysS}$
representing the outcomes for $\sysS$ and another set of operators
$\opP^{\alpha}_{\sysA}$ representing the pointer outcomes of $\sysA$. 
Presuming a given initial state $\ket{\Psi\I}_{0}$, histories are 
then given by $(\alpha_{\sysS},\alpha_{\sysA})$, with corresponding 
chain operators
\begin{equation*}
    \opC^{(\alpha_{\sysS},\alpha_{\sysA})} \equiv
	\opP^{\alpha_{\sysA}}_{\sysA}\opG_{\sysA}
	\opP^{\alpha_{\sysS}}_{\sysS}\opG_{\sysS} \,.
\end{equation*}
Further, assume that, in the given initial state, subsystems exist in 
both $\sysA$ and $\sysS$, so that
\begin{equation*}
    \opC^{(00)}\ket{\Psi\I}_{0} = 
	\opP^{0}_{\sysA}\opG_{\sysA}\opP^{0}_{\sysS}\opG_{\sysS}
	\ket{\Psi\I}_{0} = 0 \,.
\end{equation*}

Now, for $\sysA$ to be a proper measuring apparatus for $\sysS$, the 
pointer outcomes for $\sysA$ must be correlated with the outcomes of 
$\sysS$. That is,
\begin{equation} \label{eqn:E1}
    \opP^{\alpha_{\sysA}}_{\sysA}\opG_{\sysA}
    \opP^{\alpha_{\sysS}}_{\sysS} =
	\delta_{\alpha_{\sysA}\alpha_{\sysS}}
	\opP^{\alpha_{\sysS}}_{\sysA}\opG_{\sysA}
	\opP^{\alpha_{\sysS}}_{\sysS} \,,
\end{equation}
for $\alpha_{\sysA},\alpha_{\sysS} > 0$. Thus,
\begin{equation*}
    \opG_{\sysV}\ketz{\Psi\I} = \sum_{\alpha_{\sysS}>0}
	\opC^{(\alpha_{\sysS},\alpha_{\sysS})}
 	\ketz{\Psi\I} \,.
\end{equation*}

Of course, this decomposition suffers from the equivalent of a basis 
ambiguity. Let
\begin{equation*}
    \opP^{\alpha}_{\sysS} = 
	\sum_{\beta} a_{\alpha\beta} \opP^{\prime\beta}_{\sysS} \,,
\end{equation*}
for an alternate set of projection operators
$\opP^{\prime\beta}_{\sysS}$ and coefficients $a_{\alpha\beta}$ such
that
\begin{equation*}
    \sum_{\alpha} a_{\alpha\beta} = \sum_{\beta} a_{\alpha\beta} = 1 
    \\.
\end{equation*}
Then
\begin{equation*}
    \begin{split}
    \sum_{\alpha}
    \opP^{\alpha}_{\sysA}\opG_{\sysA}\opP^{\alpha}_{\sysS}
	&= \sum_{\alpha}\opP^{\alpha}_{\sysA}\opG_{\sysA}
	      \sum_{\beta} a_{\alpha\beta}\opP^{\prime\beta}_{\sysS} \\
	&= \sum_{\beta}\sum_{\alpha} a_{\alpha\beta}
	      \opP^{\alpha}_{\sysA}\opG_{\sysA}
	      \opP^{\prime\beta}_{\sysS} \\
	&= \sum_{\beta} 
	      \opP^{\prime\beta}_{\sysA}\opG_{\sysA}
	      \opP^{\prime\beta}_{\sysS} 
	      \,,
    \end{split}
\end{equation*}
where the $\opP^{\prime\beta}_{\sysA} \equiv \sum_{\alpha}
a_{\alpha\beta} \opP^{\alpha}_{\sysA}$ are an alternate set of pointer
outcomes for $\sysA$ correlated with the outcomes represented by the
$\opP^{\prime\beta}_{\sysS}$.

To resolve this, note that, if $\sysV$ really represents the entire
universe between the times $T\I$ and $T\F$, then $\sysA$ and $\sysS$
will together typically only be a small part of this. Outside of these
subsystems, there will be an \emph{environment} $\sysE = \sysV
\backslash \sysS \backslash \sysA$. There are thus \emph{three}
relevant subsystems of $\sysV$, such that
\begin{equation*}
    \opG_{\sysV}\ketz{\Psi\I}
	= \opG_{\sysE}\opG_{\sysA}\opG_{\sysS}\ketz{\Psi\I} \,.
\end{equation*}

Now, suppose that a measurement by $\sysA$ leaves a \emph{record} in 
the environment $\sysE$ and, further, that this record is independent 
of any interaction of the environment with $\sysS$. That is, there 
are outcomes of $\sysE$ represented by operators 
$\opP^{\alpha_{\sysE}}_{\sysE}$ such that
\begin{equation} \label{eqn:E2}
    \opP^{\alpha_{\sysE}}_{\sysE}\opG_{\sysE}
    \opP^{\alpha_{\sysA}}_{\sysA} =
	\delta_{\alpha_{E}\alpha_{A}}
	\opP^{\alpha_{\sysE}}_{\sysE}\opG_{\sysE}
	\opP^{\alpha_{\sysE}}_{\sysA} \,.
\end{equation}
In other words, the environment measures the apparatus. Then
\begin{equation} \label{eqn:E3}
    \opG_{\sysV}\ketz{\Psi\I} = \sum_{\alpha_{\sysS}>0}
	\opC^{(\alpha_{\sysS},\alpha_{\sysS},\alpha_{\sysS})} \,,
\end{equation}
where
\begin{equation*} 
    \opC^{(\alpha_{\sysS},\alpha_{\sysA},\alpha_{\sysE})} =
	\opP^{\alpha}_{\sysE}\opG_{\sysE}
	\opP^{\alpha}_{\sysA}\opG_{\sysA}
	\opP^{\alpha}_{\sysS}\opG_{\sysS} \,.
\end{equation*}
Such a decomposition no longer suffers from basis ambiguity. This is a
generalization to the relativistic spacetime path formalism of Zurek's
concept of \emph{einselection} \cite{zurek03a,zurek07b}.

Next consider that
\begin{equation} \label{eqn:E3a}
    \opP^{\alpha_{\sysS}}_{\sysS}\opG_{\sysS}\ketz{\Psi\I}
	= \psi^{\alpha_{\sysS}}_{\sysS}(\Psi\I)
	  \ketz{s^{\alpha_{\sysS}}(\Psi\I)} \,,
\end{equation}
where $\ketz{s^{\alpha_{\sysS}}(\Psi\I)}$ is a unit eigenstate of
$\opP^{\alpha_{\sysS}}_{\sysS}$ and $\psi^{\alpha_{\sysS}}_{\sysS}
(\Psi\I)$ is the~~magnitude~~of $\opP^{\alpha_{\sysS}}_{\sysS}
\opG_{\sysS} \ketz{\Psi\I}$. $\ketz{s^{\alpha_{\sysS}}(\Psi\I)}$
represents the outcome $\alpha_{\sysS}$ of the interaction
$\opG_{\sysS}$, given the initial state $\ket{\Psi\I}_{0}$. (If
$\psi^{\alpha_{\sysS}}_{\sysS} (\Psi\I) = 0$, then
$\ketz{s^{\alpha_{\sysS}}(\Psi\I)}$ can be chosen arbitrarily from
the eigenspace of $\opP^{\alpha_{\sysS}}_{\sysS}$.) Then, because of
\eqn{eqn:E1},
\begin{equation} \label{eqn:E4}
    \opG_{\sysA}\opP^{\alpha_{\sysS}}_{\sysS}
	= \left(
	      \sum_{\alpha_{\sysA}}\opP^{\alpha_{\sysA}}_{\sysA}
	  \right)
	  \opG^{\sysA}\opP^{\alpha_{\sysS}}_{\sysS}
	= \opP^{\alpha_{\sysS}}_{\sysA}\opG_{\sysA}
	  \opP^{\alpha_{\sysS}}_{\sysS} \,.
\end{equation}
Thus, 
\begin{equation} \label{eqn:E5}
    \opP^{\alpha_{\sysS}}_{\sysA}\opG_{\sysA}
    \opP^{\alpha_{\sysS}}_{\sysS}\opG_{\sysS} \ketz{\Psi\I}
	= \opG_{\sysA}\opP^{\alpha_{\sysS}}_{\sysS}
	  \opG_{\sysS}\ketz{\Psi\I}
	= \psi^{\alpha_{\sysS}}_{\sysS}
	  \ketz{s^{\alpha_{\sysS}},a^{\alpha_{\sysS}}} \,,
\end{equation}
where $\ketz{s^{\alpha_{\sysS}},a^{\alpha_{\sysS}}} =
\opG_{\sysA}\ketz{s^{\alpha_{\sysS}}}$ (and the explicit dependence
on $\Psi\I$ has been dropped for simplicity of notation). Because of
\eqn{eqn:E2}, a similar relationship to \eqn{eqn:E4} holds between
$\sysE$ and $\sysA$. Therefore, using this and \eqn{eqn:E5} in
\eqn{eqn:E3} gives
\begin{equation} \label{eqn:E6}
    \opG_{\sysV}\ketz{\Psi\I} =
	\sum_{\alpha_{\sysS}>0} \psi^{\alpha_{\sysS}}_{\sysS}
	\ketz{s^{\alpha_{\sysS}},a^{\alpha_{\sysS}},
	     e^{\alpha_{\sysS}}} \,,
\end{equation}
where $\ketz{s^{\alpha_{\sysS}}, a^{\alpha_{\sysS}},
e^{\alpha_{\sysS}}} = \opG_{\sysE} \ketz{s^{\alpha_{\sysS}},
a^{\alpha_{\sysS}}}$.

\Eqn{eqn:E6} is essentially the form assumed for a measurement state
in \refcite{seidewitz06b}. Each $\ketz{s^{\alpha_{\sysS}},
a^{\alpha_{\sysS}}, e^{\alpha_{\sysS}}}$ is a state of the overall
system $\sysV$ with outcome $\alpha_{\sysS}$ for $\sysS$ and
correlated outcomes for $\sysA$ and $\sysE$. And these state are
orthogonal, so \eqn{eqn:E6} certainly represents a consistent family
of decoherent histories. Thus, one clearly wants to interpret
$\sqr{\psi^{\alpha_{\sysS}}_{\sysS}}$ as the probability for
$\ketz{s^{\alpha_{\sysS}},a^{\alpha_{\sysS}},e^{\alpha_{\sysS}}}$
according to the Born rule.

The next section turns, then, to establishing the usual Born-rule
probability interpretation. Note, though, that the derivation of
\eqn{eqn:E6} is independent of this interpretation.

\section{Born's Rule} \label{sect:born}

Consider now a hypervolume $\sysV$ bounded by times $T\I$ and $T\F$
and divided into a system $\sysS$ and its environment $\sysE$, such
that
\begin{equation*}
    \opG_{\sysV} = \opG_{\sysE}\opG_{\sysS}
                 = \sum_{\alpha}\opP^{\alpha}_{\sysE}\opG_{\sysE}
		       \opP^{\alpha}_{\sysS}\opG_{\sysS} \,,
\end{equation*}
for appropriate projection operators $\opP^{\alpha}_{\sysE}$ and
$\opP^{\alpha}_{\sysS}$. Then, following a similar argument to
\sect{sect:measurements}, given an initial state $\ketz{\Psi\I}$,
\begin{equation} \label{eqn:F0}
    \opG_{\sysV}\ketz{\Psi\I} =
	\sum_{\alpha} \psi^{\alpha}_{\sysS} 
	\ketz{s^{\alpha},e^{\alpha}} \,.
\end{equation}
Suppose that the hypervolumes $\sysS$ and $\sysE$ both extend to the
final time $T\F$ and that the state of the system $\sysS$ is unchanged
by its interaction with the environment. Then it will be the case that
both
\begin{equation} \label{eqn:F1}
    \opP^{\alpha'}_{\sysS} \ketz{s^{\alpha},e^{\alpha}}
	= \delta_{\alpha'\alpha} \ketz{s^{\alpha},e^{\alpha}}
\end{equation}
and
\begin{equation} \label{eqn:F2}
    \opP^{\alpha'}_{\sysE} \ketz{s^{\alpha},e^{\alpha}}
	= \delta_{\alpha'\alpha} \ketz{s^{\alpha},e^{\alpha}} \,.
\end{equation}

Define the unitary operator
\begin{equation*}
    \opU_{\sysS} \equiv 
	\sum_{\alpha} \me^{\mi\sigma_{\alpha}}
	    \opP^{\alpha}_{\sysS} \,.
\end{equation*}
Given \eqns{eqn:F0} and \eqref{eqn:F1}, the effect of this operator is
\begin{equation*}
    \opU_{\sysS}\opG_{\sysV}\ketz{\Psi\I}
	= \sum_{\alpha} \me^{\mi\sigma_{\alpha}} \psi^{\alpha}_{\sysS}
	           \ketz{s^{\alpha},e^{\alpha}}
\end{equation*}
Because of the correlation of the environment with the system, as
reflected in \eqn{eqn:F2}, the effect of the operator $\opU_{\sysS}$
can the undone by the operator
\begin{equation*}
    \opU_{\sysE} \equiv
	 \sum_{\alpha} \me^{\mi\varepsilon_{\alpha}} 
	     \opP^{\alpha}_{\sysE} \,,
\end{equation*}
such that $\varepsilon_{j} = 2\pi\ell_{j} - \sigma_{j}$ for some 
integer $\ell_{j}$. That is,
\begin{equation*}
    \opU_{\sysE}\opU_{\sysS}\opG_{\sysV}\ketz{\Psi\I}
	= \opG_{\sysV}\ketz{\Psi\I} \,.
\end{equation*}

Now, the action of $\opU_{\sysS}$ is solely on $\sysS$. On the other 
hand, $\opU_{\sysE}$ acts solely on $\sysE$. That is, a 
transformation applied to $\sysS$ can be undone by a transformation 
applied to $\sysE$. This is a kind of symmetry that Zurek calls 
\emph{entanglement-assisted envariance}, or simply \emph{envariance} 
\cite{zurek05}. (Zurek earlier referred to this as 
``environment-assisted invariance'' \cite{zurek03a,zurek05}.)

The transformations $\opU_{\sysS}$ and $\opU_{\sysE}$ do not effect
the interaction of the system and the environment that takes place
within the overall hypervolume $\sysV$. And we have presumed that the
system and environment no longer interact outside that hypervolume.
Therefore, as argued by Zurek, we would not expect it to be possible
to undo an action on the system by an action on the causally
disconnected environment. The conclusion, then, is that any
description of the system in $\sysS$ should not depend on the phases
of the $\psi^{\alpha}_{\sysS}$, since such phases can be removed by an
action on the environment.

Since the $\opP^{\alpha_{\sysE}}_{\sysE}$ depend only on positions in
$\sysE$, while the $\opP^{\alpha_{\sysS}}_{\sysS}$ depend only on
positions in $\sysS$, $\opP^{\alpha_{\sysS}}_{\sysS}
\opP^{\alpha_{\sysE}}_{\sysE} = \opP^{\alpha_{\sysE}}_{\sysE}
\opP^{\alpha_{\sysS}}_{\sysS}$, for all $\alpha_{\sysS}$ and
$\alpha_{\sysE}$. Therefore, it is possible to conceive in general of
joint eigenstates of the $\opP^{\alpha_{\sysE}}_{\sysE}$ and
$\opP^{\alpha_{\sysS}}_{\sysS}$ with \emph{uncorrelated} outcomes such
that
\begin{equation*}
    \opP^{\alpha}_{\sysS} 
    \ketz{s^{\alpha_{\sysS}},e^{\alpha_{\sysE}}} =
	\delta_{\alpha\alpha_{\sysS}} 
	\ketz{s^{\alpha_{\sysS}},e^{\alpha_{\sysE}}}
\end{equation*}
and
\begin{equation*}
    \opP^{\alpha}_{\sysE} 
    \ketz{s^{\alpha_{\sysS}},e^{\alpha_{\sysE}}} =
	\delta_{\alpha\alpha_{\sysE}} 
	\ketz{s^{\alpha_{\sysS}},e^{\alpha_{\sysE}}}
\end{equation*}
Further, we can choose these states so that the $\ketz{s^{\alpha},
e^{\alpha}}$ with \emph{correlated} outcomes are just the states that
appear in \eqn{eqn:F0}.

Consider now the unitary operator
\begin{equation*}
    \opU^{(\beta\leftrightarrow\gamma)}_{\sysS} \equiv
	\sum_{\zeta} \left(
	    \ketz{s^{\beta},e^{\zeta}}\,\braz{s^{\gamma},e^{\zeta}} +
	    \ketz{s^{\gamma},e^{\zeta}}\,\braz{s^{\beta},e^{\zeta}} + 
	    \sum_{\alpha \neq \beta,\gamma}
		\ketz{s^{\alpha},e^{\zeta}}\,
		\braz{s^{\alpha},e^{\zeta}}
	\right) \,.
\end{equation*}
This operator has no effect on the outcomes for the environment
relative to the $\opP^{\alpha}_{\sysE}$, but it swaps the $\beta$ and
$\gamma$ outcomes for the system:
\begin{equation*}
    \opU^{(\beta\leftrightarrow\gamma)}_{\sysS}\opS_{\sysV}
    \ketz{\Psi\I} =
	\psi^{\beta}_{\sysS} \ketz{s^{\gamma},e^{\beta}} +
	\psi^{\gamma}_{\sysS} \ketz{s^{\beta},e^{\gamma}} +
	\sum_{\alpha \neq \beta,\gamma} 
	    \psi^{\alpha}_{\sysS} \ketz{s^{\alpha},e^{\alpha}} \,.
\end{equation*}

Note that the resulting state represents a different ``universe'' than
what would be expected from normal interaction based on the initial
state $\ketz{\Psi_{I}}$. It can be effectively considered to be the
result of the same basic interaction, but proceeding from a
\emph{different} initial state
\begin{equation*}
    \ketz{\Psi'\I} = 
	\opG_{\sysV}^{-1}\opU^{(\beta\leftrightarrow\gamma)}_{\sysS}
	\opG_{\sysV}\ketz{\Psi\I} \,.
\end{equation*}
So, \emph{a priori}, one cannot assume that the intrinsic properties 
of the universe represented by the swapped state will be the same as 
those of the universe represented by the original state.

However, suppose that $\psi^{\beta}_{\sysS} = \psi^{\gamma}_{\sysS}$.
Then we can apply a unitary ``counterswapping'' operator for the
environment,
\begin{equation*}
    \opU^{(\beta\leftrightarrow\gamma)}_{\sysE} \equiv
	\sum_{\zeta} \left(
	    \ketz{s^{\zeta},e^{\beta}}\,\braz{s^{\zeta},e^{\gamma}} +
	    \ketz{s^{\zeta},e^{\gamma}}\,\braz{s^{\zeta},e^{\beta}} +
	    \sum_{\alpha \neq \beta,\gamma}
		\ketz{s^{\zeta},e^{\alpha}}\,
		\braz{s^{\zeta},e^{\alpha}}
	\right) \,,
    \end{equation*}
which swaps the $\beta$ and $\gamma$ outcomes of the environment, but
leaves the outcomes of the system unchanged. Swapping first system
outcomes and then environment outcomes gives
\begin{equation*}
    \opU^{(\beta\leftrightarrow\gamma)}_{\sysS}
    \opU^{(\beta\leftrightarrow\gamma)}_{\sysE}
    \opG_{\sysV}\ketz{\Psi\I} =
	\psi^{\beta}_{\sysS} \ketz{e^{\gamma},s^{\gamma}} +
	\psi^{\gamma}_{\sysS} \ketz{e^{\beta},s^{\beta}} +
	\sum_{\alpha \neq \beta,\gamma} 
	    \psi^{\alpha}_{\sysS} \ketz{s^{\alpha},e^{\alpha}} \,.
\end{equation*}
Clearly, if $\psi^{\beta}_{\sysS} =\psi^{\gamma}_{\sysS}$,
\begin{equation*}
    \opU^{(\beta\leftrightarrow\gamma)}_{\sysS}
    \opU^{(\beta\leftrightarrow\gamma)}_{\sysE}
    \opG_{\sysV}\ketz{\Psi\I} = \opG_{\sysV}\ketz{\Psi\I} \,.
\end{equation*}
That is, a swap carried out on the system can be ``counterswapped'' 
by acting only on the environment, leaving the overall state 
unchanged: the state is envariant under swapping.

Suppose that some physical property of $\sysS$ was observably
different in the state $\opU^{(\beta\leftrightarrow\gamma)}_{\sysS}
\opG_{\sysV}\ketz{\Psi\I}$ than in $\opG_{\sysV}\ketz{\Psi\I}$. Then
this difference could be removed by acting \emph{only on the
environment} using $\opU^{(\beta\leftrightarrow\gamma)}_{\sysE}$. But
this violates the assumption that the outcome of $\sysS$ does not
depend on that of $\sysE$---that is, that information is flowing from
$\sysS$ to $\sysE$, but not vice versa.

In particular, as discussed in \refcite{seidewitz06b}, the statistics
for the results of a repeated experiment directly depend on the
probability by which the cosmological eigenstate for a given set of
outcomes is expected to be selected. Thus, if $\sysS$ includes such a
statistical measurement, any difference in the probabilities for
measurement outcomes in $\opU^{(\beta\leftrightarrow\gamma)}_{\sysS}
\opG_{\sysV}\ketz{\Psi\I}$ from $\opG_{\sysV}\ketz{\Psi\I}$ will
be physically detectable. If $\opG_{\sysV}\ketz{\Psi\I}$ is
envariant with respect to swaps, however, this should not be the case.

Therefore, we can conclude, similarly to Zurek
\cite{zurek03a,zurek05,zurek07b}, that envariant swapping cannot
effect the probabilities assigned to the system-interaction
eigenstates being swapped. That is, outcomes $\beta$ and $\gamma$ such
that $\psi^{\beta}_{\sysS} =\psi^{\gamma}_{\sysS}$ must be
\emph{equally likely}. Indeed, since we showed previously that the
phases of the $\psi^{\alpha}_{\sysS}$ can be disregarded, the real
requirement is only that $|\psi^{\beta}_{\sysS}| =
|\psi^{\gamma}_{\sysS}|$.

Given this, we can follow an approach analogous to Zurek's to obtain
Born's rule. To start, assume that the $\psi^{\alpha}_{\sysS}$ are all
rational numbers of the form
\begin{equation} \label{eqn:F3}
    \psi^{\alpha}_{\sysS} = \sqrt{m_{\alpha}/M} \,,
\end{equation}
where $\sqrt{M}$ is a common denominator of the
$\psi^{\alpha}_{\sysS}$, so that all the $m_{\alpha}$ are natural
numbers. Further, the normalization $\sum_{\alpha}
(\psi^{\alpha}_{\sysS})^{2} = 1$ gives $M = \sum_{\alpha} m_{\alpha}$.

Next, further divide the environment projection operators
$\opP^{\alpha}_{\sysE}$ into a finer-grained set
$\opP^{\alpha\beta}_{\sysE}$, such that
\begin{equation*}
    \opP^{\alpha}_{\sysE} = 
	\sum_{\beta=1}^{m_{\alpha}} \opP^{\alpha\beta}_{\sysE}
\end{equation*}
and
\begin{equation*}
    \opP^{\alpha\beta}_{\sysE}\ketz{s^{\alpha}}
	= \ketz{s^{\alpha},e^{\alpha\beta}}/\sqrt{m_{\alpha}} \,,
\end{equation*}
where the $\ketz{s^{\alpha}}$ are defined as in \eqn{eqn:E3a} and the
$\ketz{s^{\alpha}, e^{\alpha\beta}}$ are unit eigenstates. Since, in
the present formalism, all states are ultimately defined on the
infinite-dimensional space of fine-grained continuous position states,
such a discrete subdivision is always possible. Then
\begin{equation} \label{eqn:F4}
    \ketz{s^{\alpha},e^{\alpha}}
	= \opP^{\alpha}_{\sysE}\ketz{s^{\alpha}}
        = \sum_{\beta=1}^{m_{\alpha}} 
	      \ketz{s^{\alpha},e^{\alpha\beta}}/\sqrt{m_{\alpha}}\,,
\end{equation}
which respects the unit normalization of the
$\ketz{s^{\alpha},e^{\alpha}}$.

Introduce an ancillary system in a hypervolume $\sysC$ separate from
$\sysV$, but able to interact with $\sysE$ without influencing the
interaction of $\sysE$ with $\sysS$. The interaction between $\sysC$
and $\sysE$ is such that
\begin{equation} \label{eqn:F5}
    \opG_{\sysC}\opP^{\alpha\beta}_{\sysE} = 
	\delta_{\alpha\gamma}\delta_{\beta\zeta} 
	\opP^{\gamma\zeta}_{\sysC}
	\opG_{\sysC}\opP^{\alpha\beta}_{\sysE} \,,
\end{equation}
for an appropriate set of projection operators
$\opP^{\alpha\beta}_{\sysC}$ indexed parallel to the
$\opP^{\alpha\beta}_{\sysE}$. Then, using \eqns{eqn:F3},
\eqref{eqn:F4} and \eqref{eqn:F5} with \eqn{eqn:F0},
\begin{equation} \label{eqn:F6}
    \begin{split}
	\opG_{\sysC}\opG_{\sysV}\ketz{\Psi\I}
	    &= \sum_{\alpha} \sqrt{m_{\alpha}/M} 
	       \sum_{\beta=1}^{m_{\alpha}} 
	       \opG_{\sysC}\ketz{s^{\alpha},e^{\alpha\beta}}/
		   \sqrt{m_{\alpha}} \\
	    &= \sum_{\alpha\beta} \sqrt{1/M} 
	       \ketz{s^{\alpha},e^{\alpha\beta},c^{\alpha\beta}} \,,
    \end{split}
\end{equation}
where
\begin{equation*}
    \opG_{\sysC}\ket{s^{\alpha},e^{\alpha\beta}}_{0} = 
	\ketz{s^{\alpha},e^{\alpha\beta},c^{\alpha\beta}} \,,
\end{equation*}
for unit eigenstates $\ketz{s^{\alpha}, e^{\alpha\beta},
c^{\alpha\beta}}$.

The terms in \eqn{eqn:F6} all now have equal coefficients, so we take
the corresponding states to all be equally likely. Since there a total
of $M$ terms, the probability of any one of the states is $1/M$.
Further, since, for each $\alpha$, $m_{\alpha}$ of the overall
system/environment/ancilla states correspond to the system outcome
$\alpha$, the probability for this outcome is
\begin{equation*}
    p_{\alpha} = m_{\alpha}/M = \sqr{\psi^{\alpha}_{\sysS}} \,,
\end{equation*}
which is just Born's law. By continuity, the same conclusion can be
extended to all real $\psi^{\alpha}_{\sysS}$. (Note that the
derivation here also assumes the additivity of probabilities, but it
is possible to come to the same conclusion without making this
assumption \cite{zurek05}.)

Of course, this argument only establishes the Born rule for the
overall states $\ket{s^{\alpha}, e^{\alpha}}_{0}$. But
\refcite{seidewitz06b} establishes that, if Born's rule holds for such
states, then it follows that the statistics of repeated measurement
experiments would be expected to also follow this rule.

\section{Schr\"odinger's Cat} \label{sect:cat}

It is instructive to use the formalism of subsystems developed here to
analyze the classic example of macroscopic entanglement:
Schr\"odinger's Cat. The Schr\"odinger's Cat thought experiment can be
divided into five subsystems:
\begin{itemize}
    \item  $\sysR$, a radioactive atom, with projection operators 
    $\opP_{\sysR}\yes$ and $\opP_{\sysR}\no$ indicating that it has 
    or has not decayed.

    \item  $\sysD$, a detector/poison gas apparatus, with projection 
    operators $\opP_{\sysD}\yes$ and $\opP_{\sysD}\no$ indicating 
    that a decay product has been detected, with a consequent release 
    of poison gas, or not.

    \item $\sysC$, the cat, with projection operators
    $\opP_{\sysC}\alive$ and $\opP_{\sysC}\dead$ indicating that the
    cat is alive or dead.

    \item $\sysB$, the box (consisting of just the bounding container
    but not its interior), with projection operators
    $\opP_{\sysB}\open$ and $\opP_{\sysB}\closed$ indicating that the 
    box is open or closed.

    \item  $\sysE$, the environment with projection operators 
    $\opP_{\sysE}\closed$, $\opP_{\sysE}\alive$ and 
    $\opP_{\sysE}\dead$ indicating that either that the box is closed 
    or it is open and the cat is alive or dead.
\end{itemize}
The experiment is presumed to have a finite duration, so that, as
before, the complete hypervolume $\sysV = \sysE \cup \sysB \cup \sysC 
\cup \sysD \cup \sysR$ has both upper and lower time bounds.

Let $\ketz{\Psi_{I}}$ be an initial state in which the box already
exists, with the cat, atom and detector sealed inside it. That is,
\begin{equation*}
    \opP_{\sysR}\no \ketz{\Psi_{I}}
	= \opP_{\sysD}\no \ketz{\Psi_{I}}
	= \opP_{\sysC}\alive \ketz{\Psi_{I}}
	= \opP_{\sysB}\closed \ketz{\Psi_{I}}
	= \ketz{\Psi_{I}} \,.
\end{equation*}
Consider first the interior of the box, consisting of $\sysI = \sysR 
\cup \sysD \cup \sysC$. Clearly,
\begin{equation} \label{eqn:G1}
    \begin{aligned}
	\opG_{\sysI}
	    &= \opP_{\sysC}\alive \opG_{\sysC} \opP_{\sysD}\no 
	       \opG_{\sysD} \opP_{\sysR}\no \opG_{\sysR} 
	       \ketz{\Psi_{I}}
	    + \opP_{\sysC}\dead \opG_{\sysC} \opP_{\sysD}\yes 
	      \opG_{\sysD} \opP_{\sysR}\yes \opG_{\sysR} 
	      \ketz{\Psi_{I}} \\
	    &= \psi_{\sysR}\no \ketz{r\no, d\no, c\alive}
	     + \psi_{\sysR}\yes \ketz{r\yes, d\yes, c\dead} \,.
    \end{aligned}
\end{equation}

The key issue in the Schr\"odinger's Cat scenario is, of course,
whether opening the box has any relevance to the state of the interior
of the box (i.e., by ``collapsing the wave function''). Initially, the
box is closed. However, at some time during the course of the
experiment, the box may be opened, presumably by an experimenter who
is part of the environment. But, since the experimenter cannot see
inside the box while it is closed, opening the box is done with
\emph{no knowledge} of what has happened within the interior $\sysI$
of the box. If we take the interactions necessary to open the box to
be captured by $\opG_{\sysB}$, then this must commute with
$\opG_{\sysI}$ determined above.

Now, opening the box is a macroscopic, classical act which can be
presumed to either happen (with probability 1) or not. Whether the 
box is opened can thus be considered to be fully determined by the 
initial state, which includes the intention of the 
experimenter whether to open the box or not. Suppose in the initial 
state $\ketz{\Psi\I}$ the experimenter does, in fact, open the box 
some time within the hypervolume $\sysB \cup \sysI$. Then
\begin{equation*}
    \opG_{\sysB} \opG_{\sysI} \ketz{\Psi\I} =
	\opG_{\sysI} \opG_{\sysB} \ketz{\Psi\I} =
	\opG_{\sysI} \opP_{\sysB}\open \opG_{\sysB} \ketz{\Psi\I} =
	\opP_{\sysB}\open \opG_{\sysB} \opG_{\sysI} \ketz{\Psi\I} \,.
\end{equation*}

Once the box is open, the interior of the box can interact with the
environment and it becomes known in the environment whether the cat is
alive or dead in the interior of the box. So
\begin{equation*}
    \begin{split}
    \opG_{\sysV} \ketz{\Psi\I} 
	& = \opG_{\sysE} \opG_{\sysB} \opG_{\sysI} \ketz{\Psi\I} = 
	    \opG_{\sysE} \opP_{\sysB}\open \opG_{\sysB} \opG_{\sysI}
		\ketz{\Psi\I} \\
	& = \opP_{\sysE}\alive \opG_{\sysE} \opP_{\sysB}\open  
		\opG_{\sysB} \opP_{\sysC}\alive \opG_{\sysI} 
		\ketz{\Psi\I} + 
		\opP_{\sysE}\dead \opG_{\sysE} \opP_{\sysB}\open  
		\opG_{\sysB} \opP_{\sysC}\dead \opG_{\sysI} 
		\ketz{\Psi\I} \,.
    \end{split}
\end{equation*}
Thus, using \eqn{eqn:G1},
\begin{equation*}
    \begin{aligned}
	\opG_{\sysV} \ketz{\Psi\I}
	    &= \opP_{\sysE}\alive \opG_{\sysE} \opP_{\sysB}\open
	       \opG_{\sysB}\psi_{\sysR}\no 
	       \ketz{r\no, d\no, c\alive} 
	    + \opP_{\sysE}\dead \opG_{\sysE} \opP_{\sysB}\open
	      \opG_{\sysB}\psi_{\sysR}\yes 
	      \ketz{r\yes, d\yes, c\dead} \\
	    &= \psi_{\sysR}\no 
	       \ketz{r\no, d\no, c\alive, b\open, e\alive} 
	    + \psi_{\sysR}\yes 
	      \ketz{r\yes, d\yes, c\dead, b\open, e\dead} \,,
    \end{aligned}
\end{equation*}
where the environment records whether the cat is alive or dead.

However, now assume a different initial state $\ketz{\Psi'\I}$ that is
the same as $\ketz{\Psi\I}$ except that it does \emph{not} lead to the
experimenter opening the box during the time period covered by
$\sysV$. In this case
\begin{equation*}
    \opG_{\sysB} \opG_{\sysI} \ketz{\Psi'\I} =
	\opP_{\sysB}\closed \opG_{\sysB} \opG_{\sysI} \ketz{\Psi'\I} \,.
\end{equation*}
and, with the box closed, the interior cannot interact with the
environment:
\begin{equation*}
    \opG_{\sysV} \ketz{\Psi'\I} =
	\opG_{\sysE} \opG_{\sysB} \opG_{\sysI} \ketz{\Psi'\I} =
	\opP_{\sysE}\closed \opG_{\sysE} \opP_{\sysB}\closed
	    \opG_{\sysB} \opG_{\sysI} \ketz{\Psi'\I}
\end{equation*}
The change in initial state does not effect what happens in the
interior of the box, so, using \eqn{eqn:G1} again,
\begin{equation} \label{eqn:G2}
    \begin{aligned}
	\opG_{\sysV} \ketz{\Psi'\I}
	    &= \opP_{\sysE}\closed \opG_{\sysE} \opP_{\sysB}\closed  
	      \opG_{\sysB} (\opP_{\sysC}\alive + \opP_{\sysC}\dead)
	      \opG_{\sysI} \ketz{\Psi'\I} \\
	    &= \opP_{\sysE}\closed \opG_{\sysE} \opP_{\sysB}\closed
	       (\psi_{\sysR}\no \ketz{r\no, d\no, c\alive} 
	          + \psi_{\sysR}\yes \ketz{r\yes, d\yes, c\dead}
	       ) \\
	    &= \psi_{\sysR}\no 
	       \ketz{r\no, d\no, c\alive, b\closed, e\closed} 
	     + \psi_{\sysR}\yes 
	       \ketz{r\yes, d\yes, c\dead, b\closed, e\closed} \,.
    \end{aligned}
\end{equation}
The environment is now \emph{not} correlated with the alternatives
inside the box. Nevertheless there are still two orthogonal
eigenstates, representing alternative decoherent histories of the full
system, in one of which the cat is alive and in the other of which the
cat is dead. There is no alternative in which \emph{only} the cat is
in a superposition of alive and dead.

As discussed in \cite{seidewitz06b}, we can consider the state of our
actual universe to be one or the other of the alternatives in
\eqn{eqn:G2}, selected with probabilities given by
$\sqr{\psi_{\sysR}\no}$ and $\sqr{\psi_{\sysR}\yes}$. But, even though
one or the other alternative may be chosen as ``the'' state of the
universe---and the cat certainly knows which one it is!---if the box
is closed, this information is simply unavailable to the environment
outside the box. The outcome for $\sysE$ is thus the same regardless
of what happens inside the box.

Note that the above analysis is not changed if we presume that the
intent to open the box is formulated in the brain of the experimenter
sometime after the initiation of the experiment. Or if the
experimenter is replaced with, say, a device that may randomly open
the box during the run of the experiment. In all cases, by the end of
the experimental period, the box will be either open or still closed.

Whether the box is opened or remains closed, this example illustrates
how a microscopic quantum event with orthogonal outcomes can be
amplified to determine orthogonal eigenstates for an entire
macroscopic system and its environment---and, conceptually, the entire
universe. These orthogonal eigenstates represent a consistent set of
alternative histories, in one of which the cat is alive and in the
other of which the cat is dead. The alternatives of the cat being dead
and alive are thus clear and classical. Indeed, the composite
subsystem in the interior $\sysI$ of the box is already sufficient to
provide the necessary decoherence of alternatives, regardless of
whether this information can get outside of the box to its external
environment.

\section{Concluding Remarks} \label{sect:conclusion}

As with any derivation, Zurek's derivation of Born's rule is based on
a number of basic assumptions, as nicely elucidated by Schlosshaur and
Fine \cite{schlosshauer05} and further addressed by Zurek himself
\cite{zurek07b}. These assumptions condition the interpretation of the
non-relativistic formalism used by Zurek, where two entangled
systems are represented as evolving into a Schmidt state in which the
individual states of the two systems are correlated. Clearly, similar
assumptions also underlie the approach I have used here---but the
relativistic, spacetime formalism provides a rather interesting new
viewpoint on them.

The fundamental difference is that the state $G_{\sysV}
\ketz{\Phi_{I}}$ is not a Schmidt state of a cross-product Hilbert
space for systems $\sysS$ and $\sysE$ but, rather, a superposition of
joint eigenstates $\ketz{s^{\alpha},e^{\alpha}}$ of correlated
outcomes $s^{\alpha}$ and $e^{\alpha}$ for the two systems. Therefore,
it does not really make sense to speak of \emph{separate}
probabilities for the outcomes $s^{\alpha}$ and $e^{\alpha}$. There is
only the probability of whether the actual universe is a specific
joint eigenstate of these outcomes or not. The correlation of the
outcomes for $\sysS$ and $\sysE$ for any such eigenstate is completely
determined by the initial state $\ketz{\Phi_{I}}$ and the allowed
interactions within $\sysV = \sysS \cup \sysE$.

As discussed in \sect{sect:born}, the effect of a system-outcome swap
operator is to transform one state of the universe into another. The
new state can be considered as having a different effective initial
state, starting from which, interactions in the system result in
swapped outcomes. Assuming the swapped outcomes have coefficients with
the same absolute values, counterswapping the corresponding
environment outcomes then results, envariantly, in the original state.

The key assumption that Schlosshauer and Fine find most troubling is
that the probability for the system outcomes in the swapped state
remain unchanged when the counterswapping operation is applied. From
the present point of view, this assumption means the probabilities of
system outcomes in a universe based on the new effective initial state
resulting from the swap operation should be the same as the
probabilities of the system outcomes of the universe based on the
original initial state. 

However, these probabilities have physically~~observable~~consequences
with\-in each of the respective possible universes. But the fact that
the eigenstate representing one universe can be transformed into the
state of the other by applying an operator that \emph{effects only the
environment} would indicate that the physically observable properties
of the system should be the same in both universes. The specific
environment outcomes with which the system outcomes are correlated are
largely arbitrary (at least when the system outcome coefficients have
equal absolute values). They are a result of the reaction of the
environment to the system based on the initial state of the
environment, not an intrinsic property of the system.

A similar statement to the above could, of course, be made about the
probabilities of environment outcomes, since the envariance argument
is symmetrical between the system and the environment. However, there
is a deeper assumption that distinguishes the system from its
environment, which comes out clearly in the formalism: in
\eqn{eqn:F0}, it is the initial interaction of the \emph{system} with
the initial state that determines the coefficients
$\psi^{\alpha}_{\sysS}$ and the decoherence of the cosmological state
into a superposition of eigenstates of system outcomes. The further
interaction of the environment with the system simply acts to
correlate the environment with those already established eigenstates
(per the discussion on \eqn{eqn:E5} and following).

The basic assumption is that information flows \emph{from} the system
\emph{to} its environment, not vice versa. This assumption is captured
in \eqn{eqn:D4b} which states that the outcomes of interest for each
subsystem depend only on particles outgoing ``to the left'' from each
subsystem interaction operator in \eqn{eqn:D4a}. As a result, we have
been able to conveniently order subsystem interaction right to left
(e.g, $G_{\sysE} G_{\sysS}$ in \sect{sect:born}, $G_{\sysE} G_{\sysA}
G_{\sysS}$ in \sect{sect:measurements} and $G_{\sysE} G_{\sysB}
G_{\sysC} G_{\sysD} G_{\sysR}$ in \sect{sect:cat}) such that a
subsystem is affected by the outcomes of subsystems to its right, but
not by those to its left. That is, information effectively flows from
right to left.

This conception is directly related to Zurek's observation regarding
einselection on ``the direction of information flow in decoherence,
from the decohering apparatus and to the environment\ldots''
\cite{zurek03a}. This is opposite to the information flow of
\emph{noise}. In an idealized measurement situation, the desired
information flow is that necessary for decoherence and the noise
effect of the environment on the apparatus is ignored.

With the convention of \eqn{eqn:D4b}, the desired (``right to left'')
information flow is carried by particles passing from one subsystem to
the other on normal particle paths. In contrast, the undesired noise
flows ``backwards'' (``left to right''), carried by particles along
\emph{reverse} particle paths. In both cases, however, the flow of
information is along spacetime paths in the direction of increase in
the path evolution parameter, regardless of whether this is forward or
backward in time.

This interesting connection between particle propagation along
spacetime paths and the flow of information is a promising topic for
future exploration.

\appendix

\section{Spacetime Path Formalism} \label{sect:path}

This appendix summarizes the full spacetime path formalism that is
developed in detail in \refcite{seidewitz06a}. This full formalism
provides a more rigorous underpinning for the more familiar quantum
field theoretic approach described in \sect{sect:formalism}. The
formalism presented here can also be extended to particles of non-zero
spin \cite{seidewitz09}, but, for simplicity, this will not explicitly
be considered here, since the introduction of spin indices does not
fundamentally affect the points to be made in this paper. However,
note that the introduction of reverse particles in
\sect{sect:formalism} is an extension to the formalism presented in
\refcite{seidewitz06a} that is necessary to fully reproduce the
results of \sect{sect:measurements} and \sect{sect:born}.

A \emph{spacetime path} is specified by four functions $\qmul$, for
$\mu = 0, 1, 2, 3$, of a \emph{path parameter} $\lambda$. Note that
such a path is not constrained to be timelike or even to maintain any
particular direction in time. The only requirement is that it must be
continuous. And, while there is no \emph{a priori} requirement for the
paths to be differentiable, we can, as usual, treat them as
differentiable within the context of a path integral (see the
discussion in \refcite{seidewitz06a}.)

It is well known that a spacetime path integral of the form
\begin{equation} \label{eqn:AA1}
    \prop = \eta \int_{\lambdaz}^{\infty} \dif \lambda_{1}\, \intDfour q\,
            \delta^{4}(q(\lambda_{1}) - x) 
	    \delta^{4}(q(\lambdaz) - \xz)
            \exp\left( 
                \mi \int^{\lambda_{1}}_{\lambdaz} \dl L(\qdotsq(\lambda))
            \right) \,,
\end{equation}
for an appropriate normalization constant $\eta$ and the Lagrangian
function
\begin{equation*}
    L(\qdotsq) = \frac{1}{4}\qdotsq - m^{2} \,,
\end{equation*}
gives the free-particle Feynman propagator
\cite{feynman50,teitelboim82,halliwell01b,seidewitz06a}. In the path
integral above, the notation $\Dfour q$ indicates that the integral is
over the four functions $\qmul$ and the delta functions constrain the
starting and ending points of the paths integrated over. (See also
\cite{seidewitz06a} for a justification of \eqn{eqn:AA1} from a small
number of physically motivated postulates.)

Consider, however, that \eqn{eqn:AA1} can be written
\begin{equation*}
    \prop = \int_{\lambdaz}^{\infty} \dif \lambda_{1}\, 
            \kersym(x-\xz; \lambda_{1}-\lambdaz) \,,
\end{equation*}
where
\begin{equation} \label{eqn:AA2}
    \kersym(x-\xz; \lambda_{1}-\lambdaz) \equiv
        \eta \intDfour q\,
            \delta^{4}(q(\lambda_{1}) - x) 
	    \delta^{4}(q(\lambdaz) - \xz)
            \exp\left( 
                \mi \int^{\lambda_{1}}_{\lambdaz} \dl L(\qdotsq(\lambda))
            \right) \,.
\end{equation}
The value $\lambda-\lambdaz$ in $\kerneld$ can be thought of as fixing
a specific \emph{intrinsic length} for the paths being integrated
over. \Eqn{eqn:AA2} now has a similar path integral form as the usual
non-relativistic \emph{propagation kernel} \cite{feynman48,feynman65},
except with paths parametrized by $\lambda$ rather than time. We can,
therefore, use the relativistic kernel of \eqn{eqn:AA2} to define
parametrized wave function in a similar fashion to the
non-relativistic case:
\begin{equation} \label{eqn:AA3}
    \psixl = \intfour \xz\, \kerneld \psixlz \,.
\end{equation}
These wave functions are parametrized probability amplitude functions
in the sense first defined by Stueckelberg
\cite{stueckelberg41,stueckelberg42}. In this sense, the $\psixl$
represent the probability amplitude for a particle to reach position
$x$ at the point along its path with parameter value $\lambda$. (For
other related approaches using an invariant ``fifth parameter'',
though not necessarily a path
evolution parameter, see \refcites{fock37,nambu50,schwinger51,%
morette51,cooke68,horwitz73,collins78,fanchi78,piron78,fanchi83,fanchi93}.)

The functions defined in \eqn{eqn:AA3} form a Hilbert space over four
dimensional spacetime, parametrized by $\lambda$, in the same way that
traditional non-relativistic wave functions form a Hilbert space over
three dimensional space, parametrized by time. We can therefore define
a consistent family of \emph{position state} bases $\ketxl$, such that
\begin{equation} \label{eqn:AA5}
    \psixl = \innerxlpsi \,,
\end{equation}
given a single Hilbert space state vector $\ketpsi$. These position
states are normalized such that
\begin{equation*}
    \inner{x'; \lambda}{x; \lambda} = \delta^{4}(x' - x) \,.
\end{equation*}
for each value of $\lambda$. Further, it follows from \eqns{eqn:AA3} and
\eqref{eqn:AA5} that
\begin{equation} \label{eqn:AA6}
    \kerneld = \innerxlxlz \,.
\end{equation}
Thus, $\kerneld$ effectively defines a unitary transformation between 
the various Hilbert space bases $\ketxl$, indexed by the parameter
$\lambda$.
               
The overall state for propagation from $\xz$ to $x$ is given by the
superposition of the states for paths of all intrinsic lengths. If we
fix $\qmulz = \xmu_{0}$, then $\ketxl$ already includes all paths of
length $\lambda - \lambdaz$. Therefore, the overall state $\ketx$ for
the particle to arrive at $x$ should be given by the superposition of
the states $\ketxl$ for all $\lambda > \lambdaz$:
\begin{equation} \label{eqn:AA6a}
    \ketx \equiv \int_{\lambdaz}^{\infty} \dl\, \ketxl \,.
\end{equation}
Then, using \eqn{eqn:AA6},
\begin{equation*}
    \innerxxlz
           = \int_{\lambdaz}^{\infty} \dl\, \kerneld 
           = \int_{0}^{\infty} \dl\, \kersym(x-\xz; \lambda)
           = \prop \,.
\end{equation*}

Since $\kerneld$ only depends on the difference $\lambda - \lambdaz$,
the actual starting value $\lambdaz$ of the path parameter can be
shifted arbitrarily. (This can be viewed as a gauge invariance of the
path parameter $\lambda$ \cite{seidewitz06a,teitelboim82}.)
Nevertheless, it is convenient to consistently denote the starting
value for $\lambda$ as $\lambdaz$. The position states $\ketlz{x}$ can
then be identified with the states denoted $\ketxz$ in
\sect{sect:formalism}, with the states denoted $\ketx$ there being 
the same as those defined in \eqn{eqn:AA6a}.

The position states $\ketx$ as defined above make no distinction
based on the time-direction of propagation of particles. Normally,
particles are considered to propagate \emph{from} the past \emph{to}
the future. Therefore, we can define normal particle states $\ketax$
such that
\begin{equation} \label{eqn:AB0}
    \innerxaxlz = \thetaax \prop \,,
\end{equation}
On the other hand, \emph{antiparticles} may be considered to propagate
from the \emph{future} into the \emph{past}
\cite{stueckelberg41,stueckelberg42,feynman49}. Therefore,
antiparticle states $\ketrx$ are such that
\begin{equation} \label{eqn:AB0a}
    \innerxrxlz = \thetarx \prop \,.
\end{equation}

Note that the states $\ketarx$ defined here differ from the
definitions of the similarly notated states in \sect{sect:formalism}
in that the Heaviside theta functions are included in the definitions
of the states in \eqns{eqn:AB0} and \eqref{eqn:AB0a} but not in the
definitions in \sect{sect:formalism}. This means that the states
$\ketarx$ defined here are not actually on-shell, but, on the other
hand, they clearly capture the fact that particles propagate only into
the future and antiparticles propagate only into the past.
Nevertheless, as noted in \sect{sect:formalism}, we can recover
on-shell states by going to the infinite-time limit.

In taking the infinite-time limit of a spacetime path, one cannot
expect to hold the 3-position of the path end point constant. For a
free particle, though, it is reasonable to take the particle
\emph{3-momentum} as being fixed. In \refcite{seidewitz06a} it is
shown that, at the time limit of infinity (for particles) or negative
infinity (for antiparticles), such 3-momentum states do indeed become
on-shell. Thus, the momentum shell constraint is not imposed
arbitrarily but, rather, is a natural consequence of the infinite-time
limit for free particles---but only holds approximately, otherwise.

For the purposes of this paper, the 4-dimensional position states
$\ketxlz$ (or $\ketxz$, as they are denoted in the main body) are more
useful then the on-shell particle and antiparticle states. It is, of
course, straightforward to construct corresponding momentum states:
\begin{equation*}
    \ketplz \equiv (2\pi)^{-2} \intfour x\, \me^{\mi p \cdot x} 
                                           \ketlz{x} \,.
\end{equation*}
But such states are inherently off shell, with no restriction on the
value of the energy $p^{0}$ relative to the 3-momentum $\threep$.
Nevertheless, keep in mind that in any scattering-like interaction
process (as, e.g., captured in the interaction operator $\opG$ defined
in \sect{sect:formalism}) one can consider incoming and outgoing
particles to be on-shell sufficiently far outside the interaction area
\cite{seidewitz06a}.

Multiple particle states can be straightforwardly introduced as
members of a Fock space over the Hilbert space of position states
$\ketxl$. First, in order to allow for multiparticle states with
different types of particles, extend the position state of each
individual particle with a \emph{particle type index} $n$, such that
\begin{equation*}
    \inner{x',n';\lambda}{x,n;\lambda}
        = \delta_{n'n}\delta^{4}(x'-x) \,.
\end{equation*}
Then, construct a basis for the Fock space of multiparticle states as
sym\-me\-trized products of $N$ single particle states:
\begin{equation*}
    \ket{\xnliN}
        \equiv (N!)^{-1/2}
	\sum_{\text{perms }\Perm}
        \ket{\xni{\Perm 1};\lambda_{\Perm 1}} \cdots
        \ket{\xni{\Perm N};\lambda_{\Perm N}} \,,
\end{equation*}
where the sum is over all permutations $\Perm$ of $1, 2, \ldots, N$.
(When including Fermions, one needs to, of course, antisymmetrize 
rather than symmetrize the products \cite{seidewitz09}.)

It is then convenient to introduce a \emph{creation field} operator
$\oppsit(x,n;\lambda)$ such that
\begin{equation*}
    \oppsit(x,n;\lambda)\ket{\xnliN} 
        = \ket{x,n,\lambda;\xnliN} \,,
\end{equation*}
with the corresponding annihilation field $\oppsi(x,n;\lambda)$
having the commutation relation
\begin{equation*}
    [\oppsi(x',n';\lambda), \oppsit(x,n;\lambdaz)]
        = \delta_{n'n}\propsym(x'-x;\lambda-\lambdaz) \,.
\end{equation*}
Further, define
\begin{equation*}
    \oppsi(x,n) \equiv 
        \int_{\lambdaz}^{\infty} \dl\, \oppsi(x,n;\lambda) \,,
\end{equation*}
so that
\begin{equation*}
    [\oppsi(x',n'), \oppsit(x,n;\lambdaz)]
        = \delta^{n'}_{n}\propsym(x'-x) \,.
\end{equation*}

Identifying the field operators $\oppsi(x,n;\lambdaz)$ and
$\oppsi(x,n)$ defined here with $\oppsiz^{(n)}(x)$ and
$\oppsi^{(n)}(x)$ defined in \sect{sect:formalism} then completes the
grounding of the formalism used in the main text.

%% file: Paper_4.bbl
\begin{thebibliography}{53}
\expandafter\ifx\csname natexlab\endcsname\relax\def\natexlab#1{#1}\fi
\expandafter\ifx\csname bibnamefont\endcsname\relax
  \def\bibnamefont#1{#1}\fi
\expandafter\ifx\csname bibfnamefont\endcsname\relax
  \def\bibfnamefont#1{#1}\fi
\expandafter\ifx\csname citenamefont\endcsname\relax
  \def\citenamefont#1{#1}\fi
\expandafter\ifx\csname url\endcsname\relax
  \def\url#1{\texttt{#1}}\fi
\expandafter\ifx\csname urlprefix\endcsname\relax\def\urlprefix{URL }\fi
\providecommand{\bibinfo}[2]{#2}
\providecommand{\eprint}[2][]{\url{#2}}

\bibitem[{\citenamefont{Griffiths}(1984)}]{griffiths84}
\bibinfo{author}{\bibfnamefont{R.~B.} \bibnamefont{Griffiths}},
  \bibinfo{journal}{J. Stat.\ Phys.} \textbf{\bibinfo{volume}{36}},
  \bibinfo{pages}{219} (\bibinfo{year}{1984}).

\bibitem[{\citenamefont{Omn{\`e}s}(1988)}]{omnes88}
\bibinfo{author}{\bibfnamefont{R.}~\bibnamefont{Omn{\`e}s}},
  \bibinfo{journal}{J. Stat.\ Phys} \textbf{\bibinfo{volume}{53}},
  \bibinfo{pages}{893} (\bibinfo{year}{1988}).

\bibitem[{\citenamefont{Gell-Mann and Hartle}(1990)}]{gellmann90}
\bibinfo{author}{\bibfnamefont{M.}~\bibnamefont{Gell-Mann}} \bibnamefont{and}
  \bibinfo{author}{\bibfnamefont{J.}~\bibnamefont{Hartle}}, in
  \emph{\bibinfo{booktitle}{Complexity, Entropy and the Physics of
  Information}}, edited by
  \bibinfo{editor}{\bibfnamefont{W.}~\bibnamefont{Zurek}}
  (\bibinfo{publisher}{Addison-Wesley}, \bibinfo{address}{Reading},
  \bibinfo{year}{1990}), vol. \bibinfo{volume}{VIII} of
  \emph{\bibinfo{series}{Sante Fe Institute Studies in the Science of
  Complexity}}.

\bibitem[{\citenamefont{Griffiths}(2002{\natexlab{a}})}]{griffiths02}
\bibinfo{author}{\bibfnamefont{R.~B.} \bibnamefont{Griffiths}},
  \emph{\bibinfo{title}{Consistent Quantum Mechanics}}
  (\bibinfo{publisher}{Cambridge University Press},
  \bibinfo{address}{Cambridge}, \bibinfo{year}{2002}{\natexlab{a}}).

\bibitem[{\citenamefont{Feynman}(1948)}]{feynman48}
\bibinfo{author}{\bibfnamefont{R.~P.} \bibnamefont{Feynman}},
  \bibinfo{journal}{Rev.\ Mod.\ Phys.} \textbf{\bibinfo{volume}{20}},
  \bibinfo{pages}{367} (\bibinfo{year}{1948}).

\bibitem[{\citenamefont{Feynman and Hibbs}(1965)}]{feynman65}
\bibinfo{author}{\bibfnamefont{R.~P.} \bibnamefont{Feynman}} \bibnamefont{and}
  \bibinfo{author}{\bibfnamefont{A.~R.} \bibnamefont{Hibbs}},
  \emph{\bibinfo{title}{Quantum Mechanics and Path Integrals}}
  (\bibinfo{publisher}{McGraw Hill}, \bibinfo{address}{New York},
  \bibinfo{year}{1965}).

\bibitem[{\citenamefont{Caves}(1986)}]{caves86}
\bibinfo{author}{\bibfnamefont{C.~M.} \bibnamefont{Caves}},
  \bibinfo{journal}{Phys. Rev. D} \textbf{\bibinfo{volume}{33}},
  \bibinfo{pages}{1643} (\bibinfo{year}{1986}).

\bibitem[{\citenamefont{Hartle}(1991)}]{hartle91b}
\bibinfo{author}{\bibfnamefont{J.~B.} \bibnamefont{Hartle}},
  \bibinfo{journal}{Phys.\ Rev.\ D} \textbf{\bibinfo{volume}{44}},
  \bibinfo{pages}{3173} (\bibinfo{year}{1991}).

\bibitem[{\citenamefont{Hartle}(1993)}]{hartle92}
\bibinfo{author}{\bibfnamefont{J.~B.} \bibnamefont{Hartle}},
  \bibinfo{journal}{Vistas Astron.} \textbf{\bibinfo{volume}{37}},
  \bibinfo{pages}{569} (\bibinfo{year}{1993}),
  \bibinfo{note}{arXiv:gr-qc/9210004}.

\bibitem[{\citenamefont{Hartle}(1995)}]{hartle95}
\bibinfo{author}{\bibfnamefont{J.~B.} \bibnamefont{Hartle}}, in
  \emph{\bibinfo{booktitle}{Gravitation and Quantizations: Proceedings of the
  1992 Les Houches Summer School}}, edited by
  \bibinfo{editor}{\bibfnamefont{B.}~\bibnamefont{Julia}} \bibnamefont{and}
  \bibinfo{editor}{\bibfnamefont{J.}~\bibnamefont{Zinn-Justin}}
  (\bibinfo{publisher}{North-Holland}, \bibinfo{address}{Amsterdam},
  \bibinfo{year}{1995}), \bibinfo{note}{arXiv:gr-qc/9304006}.

\bibitem[{\citenamefont{Feynman}(1949)}]{feynman49}
\bibinfo{author}{\bibfnamefont{R.~P.} \bibnamefont{Feynman}},
  \bibinfo{journal}{Phys.\ Rev.} \textbf{\bibinfo{volume}{76}},
  \bibinfo{pages}{749} (\bibinfo{year}{1949}).

\bibitem[{\citenamefont{Feynman}(1950)}]{feynman50}
\bibinfo{author}{\bibfnamefont{R.~P.} \bibnamefont{Feynman}},
  \bibinfo{journal}{Phys.\ Rev.} \textbf{\bibinfo{volume}{80}},
  \bibinfo{pages}{440} (\bibinfo{year}{1950}).

\bibitem[{\citenamefont{Feynman}(1951)}]{feynman51}
\bibinfo{author}{\bibfnamefont{R.~P.} \bibnamefont{Feynman}},
  \bibinfo{journal}{Phys.\ Rev.} \textbf{\bibinfo{volume}{84}},
  \bibinfo{pages}{108} (\bibinfo{year}{1951}).

\bibitem[{\citenamefont{Teitelboim}(1982)}]{teitelboim82}
\bibinfo{author}{\bibfnamefont{C.}~\bibnamefont{Teitelboim}},
  \bibinfo{journal}{Phys.\ Rev.\ D} \textbf{\bibinfo{volume}{25}},
  \bibinfo{pages}{3159} (\bibinfo{year}{1982}).

\bibitem[{\citenamefont{Hartle and Hawking}(1983)}]{hartle83}
\bibinfo{author}{\bibfnamefont{J.~B.} \bibnamefont{Hartle}} \bibnamefont{and}
  \bibinfo{author}{\bibfnamefont{S.~W.} \bibnamefont{Hawking}},
  \bibinfo{journal}{Phys.\ Rev.\ D} \textbf{\bibinfo{volume}{28}},
  \bibinfo{pages}{2960} (\bibinfo{year}{1983}).

\bibitem[{\citenamefont{Hartle and Kucha\v{r}}(1986)}]{hartle86}
\bibinfo{author}{\bibfnamefont{J.~B.} \bibnamefont{Hartle}} \bibnamefont{and}
  \bibinfo{author}{\bibfnamefont{K.~V.} \bibnamefont{Kucha\v{r}}},
  \bibinfo{journal}{Phys. Rev. D} \textbf{\bibinfo{volume}{34}},
  \bibinfo{pages}{2323} (\bibinfo{year}{1986}).

\bibitem[{\citenamefont{Halliwell}(2001)}]{halliwell01a}
\bibinfo{author}{\bibfnamefont{J.~J.} \bibnamefont{Halliwell}},
  \bibinfo{journal}{Phys.\ Rev.\ D} \textbf{\bibinfo{volume}{64}},
  \bibinfo{pages}{044008} (\bibinfo{year}{2001}).

\bibitem[{\citenamefont{Halliwell and Thorwart}(2001)}]{halliwell01b}
\bibinfo{author}{\bibfnamefont{J.~J.} \bibnamefont{Halliwell}}
  \bibnamefont{and} \bibinfo{author}{\bibfnamefont{J.}~\bibnamefont{Thorwart}},
  \bibinfo{journal}{Phys.\ Rev.\ D} \textbf{\bibinfo{volume}{64}},
  \bibinfo{pages}{124018} (\bibinfo{year}{2001}).

\bibitem[{\citenamefont{Halliwell and Thorwart}(2002)}]{halliwell02}
\bibinfo{author}{\bibfnamefont{J.~J.} \bibnamefont{Halliwell}}
  \bibnamefont{and} \bibinfo{author}{\bibfnamefont{J.}~\bibnamefont{Thorwart}},
  \bibinfo{journal}{Phys.\ Rev.\ D} \textbf{\bibinfo{volume}{65}},
  \bibinfo{pages}{104009} (\bibinfo{year}{2002}).

\bibitem[{\citenamefont{Blencowe}(1991)}]{blencowe91}
\bibinfo{author}{\bibfnamefont{M.}~\bibnamefont{Blencowe}},
  \bibinfo{journal}{Ann. Phys. (N.Y.)} \textbf{\bibinfo{volume}{211}},
  \bibinfo{pages}{87} (\bibinfo{year}{1991}).

\bibitem[{\citenamefont{Isham et~al.}(1998)\citenamefont{Isham, Linden,
  Savvidou, and Schreckenberg}}]{isham98}
\bibinfo{author}{\bibfnamefont{C.~J.} \bibnamefont{Isham}},
  \bibinfo{author}{\bibfnamefont{N.}~\bibnamefont{Linden}},
  \bibinfo{author}{\bibfnamefont{K.}~\bibnamefont{Savvidou}}, \bibnamefont{and}
  \bibinfo{author}{\bibfnamefont{S.}~\bibnamefont{Schreckenberg}},
  \bibinfo{journal}{J. Math. Phys.} \textbf{\bibinfo{volume}{39}},
  \bibinfo{pages}{1818} (\bibinfo{year}{1998}).

\bibitem[{\citenamefont{Isham and Savvidou}(2001)}]{isham01}
\bibinfo{author}{\bibfnamefont{C.~J.} \bibnamefont{Isham}} \bibnamefont{and}
  \bibinfo{author}{\bibfnamefont{K.}~\bibnamefont{Savvidou}},
  \bibinfo{type}{Tech. Rep.} \bibinfo{number}{Imperial/TP/00-01/32},
  \bibinfo{institution}{Imperial College of Sciene} (\bibinfo{year}{2001}),
  \bibinfo{note}{arXiv:quant-ph/0110161}.

\bibitem[{\citenamefont{Griffiths}(2002{\natexlab{b}})}]{griffiths02a}
\bibinfo{author}{\bibfnamefont{R.~B.} \bibnamefont{Griffiths}},
  \bibinfo{journal}{Phys. Rev. A} \textbf{\bibinfo{volume}{66}},
  \bibinfo{pages}{062101} (\bibinfo{year}{2002}{\natexlab{b}}).

\bibitem[{\citenamefont{Seidewitz}(2007)}]{seidewitz06b}
\bibinfo{author}{\bibfnamefont{E.}~\bibnamefont{Seidewitz}},
  \bibinfo{journal}{Found. Phys.} \textbf{\bibinfo{volume}{37}},
  \bibinfo{pages}{572} (\bibinfo{year}{2007}),
  \bibinfo{note}{arXiv:quant-ph/0612023}.

\bibitem[{\citenamefont{Seidewitz}(2006)}]{seidewitz06a}
\bibinfo{author}{\bibfnamefont{E.}~\bibnamefont{Seidewitz}},
  \bibinfo{journal}{J. Math. Phys.} \textbf{\bibinfo{volume}{47}},
  \bibinfo{pages}{112302} (\bibinfo{year}{2006}),
  \bibinfo{note}{arXiv:quant-ph/0507115}.

\bibitem[{\citenamefont{Seidewitz}(2009)}]{seidewitz09}
\bibinfo{author}{\bibfnamefont{E.}~\bibnamefont{Seidewitz}},
  \bibinfo{journal}{Ann. Phys.} \textbf{\bibinfo{volume}{324}},
  \bibinfo{pages}{309} (\bibinfo{year}{2009}), \bibinfo{note}{arXiv:0804.3206
  [quant-ph]}.

\bibitem[{\citenamefont{Kent}(1990)}]{kent90}
\bibinfo{author}{\bibfnamefont{A.}~\bibnamefont{Kent}}, \bibinfo{journal}{Int.
  J. Mod. Phys. A} \textbf{\bibinfo{volume}{5}}, \bibinfo{pages}{1745}
  (\bibinfo{year}{1990}).

\bibitem[{\citenamefont{Squires}(1990)}]{squires90}
\bibinfo{author}{\bibfnamefont{E.~J.} \bibnamefont{Squires}},
  \bibinfo{journal}{Phys. Lett. A} \textbf{\bibinfo{volume}{145}},
  \bibinfo{pages}{67} (\bibinfo{year}{1990}).

\bibitem[{\citenamefont{Zurek}(1998)}]{zurek98}
\bibinfo{author}{\bibfnamefont{W.~H.} \bibnamefont{Zurek}},
  \bibinfo{journal}{Phil. Trans. R. Soc. Lond. A}
  \textbf{\bibinfo{volume}{356}}, \bibinfo{pages}{1793} (\bibinfo{year}{1998}).

\bibitem[{\citenamefont{Zurek}(2003{\natexlab{a}})}]{zurek03a}
\bibinfo{author}{\bibfnamefont{W.~H.} \bibnamefont{Zurek}},
  \bibinfo{journal}{Rev. Mod. Phys.} \textbf{\bibinfo{volume}{75}},
  \bibinfo{pages}{715} (\bibinfo{year}{2003}{\natexlab{a}}).

\bibitem[{\citenamefont{Zurek}(2003{\natexlab{b}})}]{zurek03b}
\bibinfo{author}{\bibfnamefont{W.~H.} \bibnamefont{Zurek}},
  \bibinfo{journal}{Phys. Rev. Lett.} \textbf{\bibinfo{volume}{90}},
  \bibinfo{pages}{120404} (\bibinfo{year}{2003}{\natexlab{b}}).

\bibitem[{\citenamefont{Zurek}(2005)}]{zurek05}
\bibinfo{author}{\bibfnamefont{W.~H.} \bibnamefont{Zurek}},
  \bibinfo{journal}{Phys. Rev. A} \textbf{\bibinfo{volume}{71}},
  \bibinfo{pages}{052105} (\bibinfo{year}{2005}).

\bibitem[{\citenamefont{Zurek}(2007{\natexlab{a}})}]{zurek07a}
\bibinfo{author}{\bibfnamefont{W.~H.} \bibnamefont{Zurek}},
  \bibinfo{journal}{Phys. Rev. A} \textbf{\bibinfo{volume}{76}},
  \bibinfo{pages}{052110} (\bibinfo{year}{2007}{\natexlab{a}}),
  \bibinfo{note}{arXiv:quant-ph/0703160}.

\bibitem[{\citenamefont{Zurek}(2007{\natexlab{b}})}]{zurek07b}
\bibinfo{author}{\bibfnamefont{W.~H.} \bibnamefont{Zurek}},
  \bibinfo{type}{Tech. Rep.} \bibinfo{number}{LAUR 07-4568},
  \bibinfo{institution}{Los Alamos National Laboratory}
  (\bibinfo{year}{2007}{\natexlab{b}}), \bibinfo{note}{arXiv:0707.2832
  [quant-ph]}.

\bibitem[{\citenamefont{Halliwell and Wallden}(2006)}]{halliwell06}
\bibinfo{author}{\bibfnamefont{J.~J.} \bibnamefont{Halliwell}}
  \bibnamefont{and} \bibinfo{author}{\bibfnamefont{P.}~\bibnamefont{Wallden}},
  \bibinfo{journal}{Phys. Rev. D} \textbf{\bibinfo{volume}{73}},
  \bibinfo{pages}{024011} (\bibinfo{year}{2006}),
  \bibinfo{note}{arXiv:quant-ph/0301117}.

\bibitem[{\citenamefont{Peskin and Schroeder}(1995)}]{peskin95}
\bibinfo{author}{\bibfnamefont{M.~E.} \bibnamefont{Peskin}} \bibnamefont{and}
  \bibinfo{author}{\bibfnamefont{D.~V.} \bibnamefont{Schroeder}},
  \emph{\bibinfo{title}{An Introduction to Quantum Field Theory}}
  (\bibinfo{publisher}{Addison-Wesley}, \bibinfo{address}{Reading,
  Massachusetts}, \bibinfo{year}{1995}).

\bibitem[{\citenamefont{Weinberg}(1995)}]{weinberg95}
\bibinfo{author}{\bibfnamefont{S.}~\bibnamefont{Weinberg}},
  \emph{\bibinfo{title}{The Quantum Theory of Fields}}, vol.
  \bibinfo{volume}{1. \emph{Foundations}} (\bibinfo{publisher}{Cambridge
  University Press}, \bibinfo{address}{Cambridge}, \bibinfo{year}{1995}).

\bibitem[{\citenamefont{Ticciati}(1999)}]{ticciati99}
\bibinfo{author}{\bibfnamefont{R.}~\bibnamefont{Ticciati}},
  \emph{\bibinfo{title}{Quantum Field Theory for Mathematicians}}
  (\bibinfo{publisher}{Cambridge University Press},
  \bibinfo{address}{Cambridge}, \bibinfo{year}{1999}).

\bibitem[{\citenamefont{Stueckelberg}(1941)}]{stueckelberg41}
\bibinfo{author}{\bibfnamefont{E.~C.~G.} \bibnamefont{Stueckelberg}},
  \bibinfo{journal}{Helv.\ Phys.\ Acta} \textbf{\bibinfo{volume}{14}},
  \bibinfo{pages}{588} (\bibinfo{year}{1941}).

\bibitem[{\citenamefont{Stueckelberg}(1942)}]{stueckelberg42}
\bibinfo{author}{\bibfnamefont{E.~C.~G.} \bibnamefont{Stueckelberg}},
  \bibinfo{journal}{Helv.\ Phys.\ Acta} \textbf{\bibinfo{volume}{15}},
  \bibinfo{pages}{23} (\bibinfo{year}{1942}).

\bibitem[{\citenamefont{Wichmann and Circhton}(1963)}]{wichmann63}
\bibinfo{author}{\bibfnamefont{E.~H.} \bibnamefont{Wichmann}} \bibnamefont{and}
  \bibinfo{author}{\bibfnamefont{J.~H.} \bibnamefont{Circhton}},
  \bibinfo{journal}{Phys. Rev.} \textbf{\bibinfo{volume}{132}},
  \bibinfo{pages}{2788} (\bibinfo{year}{1963}).

\bibitem[{\citenamefont{Schlosshauer and Fine}(2005)}]{schlosshauer05}
\bibinfo{author}{\bibfnamefont{M.}~\bibnamefont{Schlosshauer}}
  \bibnamefont{and} \bibinfo{author}{\bibfnamefont{A.}~\bibnamefont{Fine}},
  \bibinfo{journal}{Found. Phys.} \textbf{\bibinfo{volume}{35}},
  \bibinfo{pages}{197} (\bibinfo{year}{2005}).

\bibitem[{\citenamefont{Fock}(1937)}]{fock37}
\bibinfo{author}{\bibfnamefont{V.~A.} \bibnamefont{Fock}},
  \bibinfo{journal}{Physik Z. Sowjetunion} \textbf{\bibinfo{volume}{12}},
  \bibinfo{pages}{404} (\bibinfo{year}{1937}).

\bibitem[{\citenamefont{Nambu}(1950)}]{nambu50}
\bibinfo{author}{\bibfnamefont{Y.}~\bibnamefont{Nambu}},
  \bibinfo{journal}{Progr.\ Theoret.\ Phys.} \textbf{\bibinfo{volume}{5}},
  \bibinfo{pages}{82} (\bibinfo{year}{1950}).

\bibitem[{\citenamefont{Schwinger}(1951)}]{schwinger51}
\bibinfo{author}{\bibfnamefont{J.}~\bibnamefont{Schwinger}},
  \bibinfo{journal}{Phys.\ Rev.} \textbf{\bibinfo{volume}{82}},
  \bibinfo{pages}{664} (\bibinfo{year}{1951}).

\bibitem[{\citenamefont{Morette}(1951)}]{morette51}
\bibinfo{author}{\bibfnamefont{C.}~\bibnamefont{Morette}},
  \bibinfo{journal}{Phys. Rev.} \textbf{\bibinfo{volume}{81}},
  \bibinfo{pages}{848} (\bibinfo{year}{1951}).

\bibitem[{\citenamefont{Cooke}(1968)}]{cooke68}
\bibinfo{author}{\bibfnamefont{J.~H.} \bibnamefont{Cooke}},
  \bibinfo{journal}{Phys.\ Rev.} \textbf{\bibinfo{volume}{166}},
  \bibinfo{pages}{1293} (\bibinfo{year}{1968}).

\bibitem[{\citenamefont{Horwitz and Piron}(1973)}]{horwitz73}
\bibinfo{author}{\bibfnamefont{L.~P.} \bibnamefont{Horwitz}} \bibnamefont{and}
  \bibinfo{author}{\bibfnamefont{C.}~\bibnamefont{Piron}},
  \bibinfo{journal}{Helv.\ Phys.\ Acta} \textbf{\bibinfo{volume}{46}},
  \bibinfo{pages}{316} (\bibinfo{year}{1973}).

\bibitem[{\citenamefont{Collins and Fanchi}(1978)}]{collins78}
\bibinfo{author}{\bibfnamefont{R.~E.} \bibnamefont{Collins}} \bibnamefont{and}
  \bibinfo{author}{\bibfnamefont{J.~R.} \bibnamefont{Fanchi}},
  \bibinfo{journal}{Nuovo Cimento} \textbf{\bibinfo{volume}{48A}},
  \bibinfo{pages}{314} (\bibinfo{year}{1978}).

\bibitem[{\citenamefont{Fanchi and Collins}(1978)}]{fanchi78}
\bibinfo{author}{\bibfnamefont{J.~R.} \bibnamefont{Fanchi}} \bibnamefont{and}
  \bibinfo{author}{\bibfnamefont{R.~E.} \bibnamefont{Collins}},
  \bibinfo{journal}{Found.\ Phys.} \textbf{\bibinfo{volume}{8}},
  \bibinfo{pages}{851} (\bibinfo{year}{1978}).

\bibitem[{\citenamefont{Piron and Reuse}(1978)}]{piron78}
\bibinfo{author}{\bibfnamefont{C.}~\bibnamefont{Piron}} \bibnamefont{and}
  \bibinfo{author}{\bibfnamefont{F.}~\bibnamefont{Reuse}},
  \bibinfo{journal}{Helv.\ Phys.\ Acta} \textbf{\bibinfo{volume}{51}},
  \bibinfo{pages}{146} (\bibinfo{year}{1978}).

\bibitem[{\citenamefont{Fanchi and Wilson}(1983)}]{fanchi83}
\bibinfo{author}{\bibfnamefont{J.~R.} \bibnamefont{Fanchi}} \bibnamefont{and}
  \bibinfo{author}{\bibfnamefont{W.~J.} \bibnamefont{Wilson}},
  \bibinfo{journal}{Found.\ Phys.} \textbf{\bibinfo{volume}{13}},
  \bibinfo{pages}{571} (\bibinfo{year}{1983}).

\bibitem[{\citenamefont{Fanchi}(1993)}]{fanchi93}
\bibinfo{author}{\bibfnamefont{J.~R.} \bibnamefont{Fanchi}},
  \emph{\bibinfo{title}{Parametrized Relativistic Quantum Theory}}
  (\bibinfo{publisher}{Kluwer Academic}, \bibinfo{address}{Dordrecht},
  \bibinfo{year}{1993}).

\end{thebibliography}
